\documentclass[a4paper,fleqn]{cas-dc}

\usepackage[authoryear,round]{natbib}
\setcitestyle{authoryear,round,aysep={,},yysep={;}}
\usepackage{graphicx}
\usepackage{booktabs}
\usepackage{colortbl,xcolor}
\definecolor{promptRole}{RGB}{230, 126, 34}
\definecolor{promptCot}{RGB}{142, 68, 173}
\definecolor{promptGuard}{gray}{0.35}
\definecolor{promptStruct}{RGB}{0, 120, 122}
\definecolor{channelLt}{RGB}{41, 128, 185}
\definecolor{channelSt}{RGB}{52, 152, 219}
\definecolor{channelLike}{RGB}{42, 126, 2}
\definecolor{channelDis}{RGB}{192, 57, 43}
\definecolor{channelIp}{RGB}{78, 0, 255}
\definecolor{channelCfAnchor}{RGB}{0, 120, 122}
\definecolor{channelCfReject}{RGB}{192, 57, 43}
\definecolor{channelCfCond}{RGB}{142, 68, 173}
\usepackage{multirow}
\usepackage{amsmath,amssymb}
\usepackage{longtable}
\usepackage{needspace}
\usepackage{array}
\usepackage{tabularx}
\usepackage{makecell}
\usepackage{algorithm}
\usepackage{algpseudocode}
\usepackage{pifont}
\newcommand{\circled}[1]{\ding{\numexpr171+#1\relax}}
\makeatletter
\newcolumntype{P}[1]{>{\raggedright\arraybackslash}p{#1}}
\newcolumntype{Y}{>{\raggedright\arraybackslash}X}
\makeatother
\newcommand{\best}[1]{\textbf{#1}}
\newcommand{\second}[1]{\underline{#1}}
\graphicspath{{./figures/}{./}{../results/figures/}{../results/figures/ablation/}{../results/figures/optimize/}{../results/figures/bank/}}

\begin{document}
\let\WriteBookmarks\relax
\def\floatpagepagefraction{1}
\def\textpagefraction{.001}

\shorttitle{rEDMRec: Reasoning Distillation into Experience Memory}
\shortauthors{M.H. Nguyen et al.}

\title[mode=title]{rEDMRec: Distilling Large Language Model Reasoning into an Editable Experience Memory for Recommendation}

\author[1,2]{Minh Hoang Nguyen}[orcid=0009-0004-1384-3856]
\ead{24C15049@student.hcmus.edu.vn}
\credit{Methodology, Conceptualization, Writing -- original draft, Writing -- review \& editing}

\author[1,2]{Tung Le}[orcid=0000-0002-9900-7047]
\ead{lttung@fit.hcmus.edu.vn}
\credit{Supervision, Supporting, Writing -- review \& editing}

\author[1,2]{Huy Tien Nguyen}[orcid=0000-0002-9948-1048]
\cormark[1]
\ead{ntienhuy@fit.hcmus.edu.vn}
\credit{Supervision, Supporting, Conceptualization, Project administration}

\affiliation[1]{organization={Faculty of Information Technology, University of Science},
            city={Ho Chi Minh City},
            country={Vietnam}}
\affiliation[2]{organization={Vietnam National University},
            city={Ho Chi Minh City},
            country={Vietnam}}

\cortext[1]{Corresponding author}

\begin{abstract}
Large language models (LLMs) can improve recommendation quality by reasoning explicitly over user history and candidate items -- for example, extracting a user's preferences or explaining why one item fits better than another -- rather than mapping history directly to a ranked list. This reasoning, however, is expensive to repeat on every ranking request and, once produced, is typically consumed once and discarded, leaving it neither reusable across future requests nor easy to inspect or correct as user tastes drift. Our insight is that reasoning does not need to be regenerated at every call if it can instead be compressed once into a compact, structured memory that a lightweight model retrieves from. We propose rEDMRec, which distills a teacher LLM's reasoning into four typed, editable experience channels -- long-term preference, short-term context, item-perception, and counterfactual hard-negative comparisons -- maintained by an LLM memory controller that performs Add/Delete/Modify/Keep operations and refines entries via $K$-agent debate. A lightweight student LLM (3B--20B parameters) then ranks candidates purely by retrieving from this memory, without invoking the teacher again, decoupling online inference cost from reasoning depth. Across ML-1M, Amazon Beauty, and Steam and ten student backbones, rEDMRec improves HR@1 over zero-shot, few-shot, and RAG on every backbone, and over GraphRAG on most backbones (exceptions: Llama~3.1~8B and GPT~OSS~20B), with Impv up to $13.3\%$ vs.\ the second-best baseline on ML-1M, following the relative-improvement protocol of RDRec~\citep{wang2024rdrec}. Channel ablations show that short-term context is the only channel that helps consistently across capacity tiers, whereas long-term, item-perception, and counterfactual contributions are capacity-dependent (and can reverse on the strongest students); debate-based memory optimization lowers bank duplication by $7.4$ percentage points while raising downstream HR@1 by up to $+0.029$ over six optimization epochs.
\end{abstract}



\maketitle

\section{Introduction}
\label{sec:intro}

Large language models (LLMs) are increasingly used as recommenders -- via prompting, instruction tuning, collaborative-signal fusion, or generative item prediction~\citep{wu2024surveyllm4rec,lin2025howllmrec,bao2023tallrec,liao2024llara,hou2024llmrank} -- and, more recently, as explicit reasoners over user history and candidate items~\citep{wei2022cot}. An LLM can extract preferences, judge item fit, or contrast a candidate against a hard negative, then use that reasoning to guide ranking. This paper studies a concrete bottleneck that follows from that capability: once such reasoning has been produced for a user, how can it be \emph{reused} across future ranking requests, rather than regenerated from scratch every time?

Reasoning-augmented recommenders face a tension between reasoning depth and inference cost. Zero-shot and few-shot prompting, and retrieval-augmented generation (RAG) over raw interaction history~\citep{lewis2020rag}, are cheap but skip explicit preference-level reasoning, so ranking remains opaque and brittle under short histories. Closer to our setting, ReasoningRec~\citep{bismay2025reasoningrec} uses a teacher LLM to synthesize user profiles, item descriptions, and human-interpretable explanatory reasoning, then instruction-tunes a smaller model on those traces; R2Rec~\citep{r2rec} similarly builds interaction-of-thought chains and internalizes them with supervised fine-tuning and reinforcement learning; and $R^{4}$ec~\citep{gu2025r4ec} iterates actor reasoning with a reflection model that critiques and refines preference/item knowledge before feeding a recommendation backbone. These lines treat reasoning primarily as a per-request or training-time signal: competence lives in regenerated traces or model weights, cannot be updated entry-by-entry when new interactions arrive or low-quality reasoning accumulates, and revision requires another costly reasoning loop or retraining rather than a targeted memory edit. Related distillation methods compress rationales or prompts into smaller generators~\citep{wang2024rdrec,li2023pod}, yet still treat the distilled artifact as a fixed model rather than as a typed bank. The remaining challenge is therefore architectural: keep the benefit of teacher-level reasoning while making that reasoning reusable across requests and editable over time, without retraining the backbone whenever the underlying knowledge must change.

\begin{figure}[pos=t]
\centering
\includegraphics[width=\columnwidth]{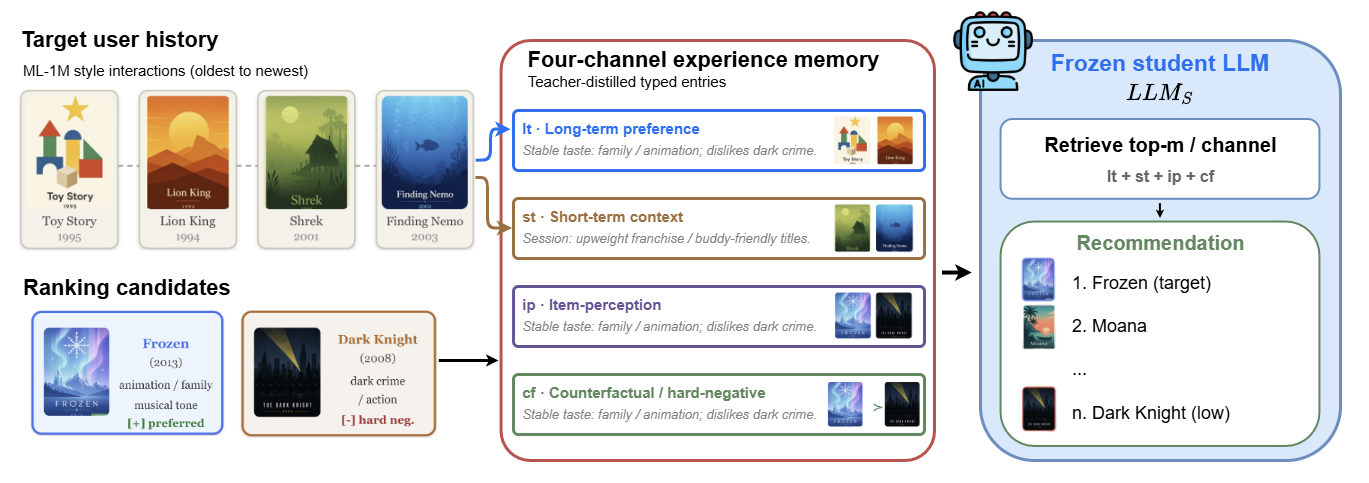}
\caption{Example of four-channel experience memory distilled from a user's history for ranking \textit{Frozen} vs.\ \textit{The Dark Knight}. Offline, teacher reasoning is stored as typed \texttt{lt}/\texttt{st}/\texttt{ip}/\texttt{cf} entries; online, a frozen student LLM retrieves those channels and produces the recommendation list without calling the teacher.}
\label{fig:teaser}
\end{figure}

As illustrated in Figure~\ref{fig:teaser}, consider a user whose history is dominated by family and animation titles (e.g., \textit{Toy Story}, \textit{The Lion King}, \textit{Shrek}, \textit{Finding Nemo}) and who must rank \textit{Frozen} against \textit{The Dark Knight}. If the recommender focuses solely on explicit item titles or undifferentiated history text, it struggles to surface the subtle connection across these interactions -- that the user consistently prefers light family entertainment over dark crime -- and has no structured place to store a hard-negative contrast that would push \textit{The Dark Knight} down when that pattern holds. In this case, the latent preference is not a single keyword but a typed bundle of signals: long-term taste (stable family/animation preference; dislike of dark crime tone), short-term session context (franchise- or buddy-friendly titles), item-perception that grounds why \textit{Frozen} matches the history while \textit{The Dark Knight} mismatches, and a counterfactual that records when the dark-action alternative would have ranked higher. Existing reasoning-augmented methods either regenerate such comparisons per request or absorb them into weights, so they cannot keep these four signals as independently retrievable and independently editable memory entries across future ranking calls.

We address this challenge with rEDMRec. The basic idea is illustrated in Figure~\ref{fig:teaser}: instead of asking an LLM to both reason and rank on every inference call, we use the LLM as a \emph{teacher reasoning generator} and distill its outputs into a structured, retrievable, and updatable Experience Memory that a lightweight student consumes at ranking time. Concretely, the teacher produces four typed experience signals -- long-term preference, short-term context, item perception, and counterfactual hard-negative comparisons -- and a distillation adapter normalizes each signal into a channel-indexed memory entry (vector store for the first three channels; hybrid vector--graph store for counterfactuals). At inference, a \emph{frozen} student (3B--20B) retrieves the top-$m$ entries per channel, composes a ranking prompt, and scores candidates without re-invoking the teacher, so accuracy gains are attributable to the memory rather than to student-parameter adaptation. A remaining difficulty is that a bank filled once from teacher extractions accumulates near-duplicates and low-specificity entries as interactions grow. Inspired by Training-Free GRPO~\citep{chen2025trainingfreegrpo}, which steers a frozen policy by maintaining an experiential knowledge library with Add/Delete/Modify/Keep operations rather than gradient updates, our second contribution is an LLM memory controller that applies the same edit operators -- optionally guided by $K$-agent debate and an arbiter -- to revise the recommendation experience bank over optimization epochs. Together, these components separate an expensive, infrequent reasoning-compression process from a cheap, frequent retrieval-and-rank process.

We evaluate rEDMRec on ML-1M, Amazon Beauty, and Steam with ten student LLMs (3B--20B), comparing against zero-shot, few-shot, RAG, and GraphRAG~\citep{edge2024graphrag}. rEDMRec improves HR@1 over zero-shot, few-shot, and RAG on every backbone, and over GraphRAG on most backbones (exceptions: Llama~3.1~8B and GPT~OSS~20B), reaching Impv $=13.3\%$ vs.\ GraphRAG on Qwen2.5~3B under the RDRec relative-improvement protocol~\citep{wang2024rdrec} (Section~\ref{sec:results-h1}). Channel ablations show that short-term context is the only consistently helpful channel across capacity tiers, whereas long-term, item-perception, and counterfactual effects are capacity-dependent (Section~\ref{sec:results-h2}). A teacher-distillation study further shows that bank duplicate rate is a leading indicator of downstream gain, bounded by student capacity. Finally, debate-based optimization lowers bank duplication by $7.4$ percentage points while raising downstream HR@1 by up to $+0.029$ over six epochs, validating the controller as a functional component of the architecture (Section~\ref{sec:results-h4}).

This paper makes the following contributions:
\begin{enumerate}
\item \textbf{An editable, channel-structured experience memory} that distills teacher LLM reasoning into four typed, independently retrievable and independently editable channels, rather than a single undifferentiated reasoning trace (Section~\ref{sec:method-memory}).
\item \textbf{A debate-based LLM memory controller} that revises the bank after each student prediction via Add/Delete/Modify/Keep operations guided by ranking reward models and $K$-agent debate, with bank-quality gains shown to propagate into ranking accuracy without updating the frozen student (Sections~\ref{sec:method-memory}, \ref{sec:method-optimize}; Section~\ref{sec:results-h4}).
\item \textbf{A comprehensive empirical study} across ten student backbones and three datasets, including channel ablations that explain when ablating a channel can \emph{improve} accuracy, a teacher-distillation study isolating teacher quality from student capacity, and a qualitative case study of entry-level edits over optimization epochs (Sections~\ref{sec:results-h1}--\ref{sec:results-qualitative}).
\end{enumerate}

\section{Related Work}
\label{sec:related}

\subsection{LLM-based Recommendation}
\label{sec:related-llm4rec}

Before LLM recommenders, neural models already encoded dual-scale user interest, review text, and cold-start evaluation -- but they stored that structure in parameters or in the corpus, not in an editable experience bank. Sequential news recommenders such as Co-NAML-LSTUR~\citep{nguyen2025co} jointly learn multi-view item encodings with long- and short-term user representations, showing that separating durable taste from recent browsing improves ranking even without an LLM. Review-based models such as RRS~\citep{nguyen2024rrs} replace ID-only collaborative filtering with deep encoders over user-written text, so item semantics enter ranking through review content rather than through a retrieved, typed memory entry. Complementary dataset work such as ViHoRec~\citep{nguyen2026vihorec, nguyen2025enhancing} shows that, on sparse Vietnamese hotel interactions with a temporal cold-start split, neighborhood methods can outperform learned latent-factor models on users with short histories. These lines motivate the same signal types that rEDMRec stores as channels -- long-term preference, short-term context, and item-level text -- yet they cannot persist teacher-generated reasoning or Add/Delete/Modify a typed entry after a ranking failure. rEDMRec keeps that factorization, but as a non-parametric bank that a frozen student retrieves from, rather than as a neural user/item tower.

Recent surveys organize LLM recommenders into prompting, tuning, collaborative fusion, generative recommendation, and system-enhancement paradigms~\citep{wu2024surveyllm4rec,lin2025howllmrec,wang2024nextgenllmrec,liu2025llmenhancedsurvey,li2024generativerecsurvey}. Prompting and instruction-following methods cast ranking as natural-language generation or zero-shot ordering~\citep{gao2023chatrec,zhang2023instructrec,hou2024llmrank,yue2023llamarec,lyu2024llmrecprompt}, while alignment frameworks such as TALLRec and LLaRA adapt LLMs to recommendation with efficient fine-tuning~\citep{bao2023tallrec,liao2024llara}. A parallel line injects collaborative or ID structure into the language space -- CoLLM, BinLLM, TokenRec, and related models encode interaction signals as text-like or tokenized representations~\citep{zhang2025collm,zhang2024binllm,qu2024tokenrec} -- and enhancement methods use LLMs for graph augmentation, tool use, or query generation around a classical backbone~\citep{wei2024llmrec,zhao2024toolrec,han2025querec}. Across these paradigms, any intermediate reasoning (when present) remains either ephemeral prompt context or knowledge absorbed into parameters. rEDMRec instead materializes teacher reasoning as a non-parametric, typed memory that survives across sessions and can be revised without re-tuning the student.

\subsection{Reasoning, Distillation, and Preference Utilization}
\label{sec:related-reasoning}

A narrower line makes LLM reasoning itself the object of design. ReasoningRec~\citep{bismay2025reasoningrec} synthesizes user profiles, item descriptions, and explanatory rationales with a teacher LLM, then instruction-tunes a smaller model for both prediction and human-interpretable explanation -- an extraction-then-fine-tune pipeline rather than a durable memory architecture. R2Rec~\citep{r2rec} samples interaction chains, builds interaction-of-thought traces, and internalizes them with SFT and RL; LatentR$^3$~\citep{zhang2025latentr3} similarly reinforces latent reasoning inside the model, while SPRec~\citep{gao2025sprec} uses self-play to debias generative recommenders. $R^{4}$ec~\citep{gu2025r4ec} pushes toward System-2 deliberation by pairing an actor that proposes preference/item knowledge with a reflection model that judges and triggers refinement until the knowledge is deemed rational, then injects the refined text into a recommendation backbone -- still a per-case reasoning loop rather than a persistent, editable experience bank. Distillation work such as RDRec, POD, and LEADER compresses rationales or teacher signals into smaller recommendation models~\citep{wang2024rdrec,li2023pod,liu2024leader}. Retrieval baselines (RAG, GraphRAG) ground ranking in raw history or graph summaries without teacher-compressed experience~\citep{lewis2020rag,edge2024graphrag}. Relative to these methods, rEDMRec neither stops at one-shot feature utilization nor freezes reasoning into weights: it distills reasoning into a channel-typed bank with entry-level Add/Delete/Modify/Keep.

\subsection{Memory-Augmented Agents and Recommendation Memory}
\label{sec:related-memory}

External memory lets LLM agents operate beyond a single context window. MemGPT~\citep{packer2023memgpt} pages context like an OS virtual-memory manager; Generative Agents~\citep{park2023generativeagents} maintain a retrieve-and-reflect memory stream for long-horizon behavior. Closest to our controller design, Training-Free GRPO~\citep{chen2025trainingfreegrpo} improves a frozen LLM by iteratively distilling experiential knowledge into a non-parametric library updated with Add/Delete/Modify/Keep operations -- a training-free alternative to gradient-based GRPO. In recommendation, AutoMR retrieves stored experiences for generative ranking~\citep{wang2025leveraging}, long-term planners model durable taste~\citep{shi2024large}, and related work motivates separating long- versus short-term interest~\citep{zheng2024llmtrsr,zhang2024lsidn}, item-level semantics~\citep{ren2024rlmrec,zhang2026token2item}, hard-negative contrast~\citep{song2026llmhni,li2026ilrec}, and user-controllable profiles~\citep{wozniak2025improving}. These designs retrieve history, profiles, or generic agent traces; they are not organized as teacher-distilled, four-channel recommendation experience with debate-driven bank maintenance. rEDMRec adopts the Training-Free GRPO principle of editing an external experience library instead of model weights, but specializes the schema to recommendation signal types and pairs each channel with a matching store (vector vs.\ hybrid graph--vector).

\subsection{Multi-Agent Debate and Self-Refinement}
\label{sec:related-debate}

Iterative critique improves LLM outputs without additional supervised labels. Self-Refine~\citep{madaan2023selfrefine} has a model revise its own text from self-feedback; multiagent debate~\citep{du2023multiagentdebate} improves factuality by having several instances propose and critique answers; and in recommendation, $R^{4}$ec~\citep{gu2025r4ec} couples actor and reflection models to refine preference/item knowledge before backbone prediction. These lines evaluate a single artifact -- one answer, one document, or one knowledge string per case -- and do not measure how repeated refinement of a growing collection affects duplicate accumulation or retrieval quality at bank scale. rEDMRec applies critique-and-revise to the experience bank: $K$ debating personas critique a case, an arbiter synthesizes revisions, and the controller commits Add/Delete/Modify/Keep operations, with bank-level duplicate rate and specificity linked to downstream HR@1 (Section~\ref{sec:results-h4}).

\section{Method}
\label{sec:method}

We formalize next-item ranking as conditional generation (Section~\ref{sec:method-formulation}), restate the dominant prompting and retrieval paradigms in the same notation to make the technical gap precise (Section~\ref{sec:method-prelim}), then specify rEDMRec's components as LLM operators over a typed non-parametric memory (Sections~\ref{sec:method-adapter}--\ref{sec:method-optimize}). Table~\ref{tbl:notation} collects the notation used throughout.

\begin{table}[pos=t]
\caption{Notation used in Section~\ref{sec:method}.}
\label{tbl:notation}
\footnotesize
\centering
\begin{tabularx}{\columnwidth}{@{}P{0.40\columnwidth}Y@{}}
\toprule
Symbol & Meaning \\
\midrule
$u$, $H_u$ & user; chronological interaction history \\
$\mathcal{I}$, $M(i)$ & item catalog; metadata text of item $i$ \\
$C_u\subset\mathcal I$, $c$ & 20-item candidate set; a candidate in $C_u$ \\
$i^+$, $i^-$, $\hat{i}$ & positive target; hard-negative contrast; predicted item \\
$x_u$, $D_K$ & ranking prompt; $K$ in-context demonstrations (few-shot) \\
$\theta$, $P_S$ & generic LLM parameters; student likelihood under $\mathrm{LLM}_S$ \\
$\mathrm{Enc}(\cdot)$, $\mathrm{sim}$ & shared text encoder; cosine similarity \\
$\mathbf{q}_u$, $d$ & query embedding $\mathrm{Enc}(\mathrm{prompt}(u,C_u))$; embedding dim. \\
$\mathcal K$, $k$ & channel set $\{\mathrm{lt},\mathrm{st},\mathrm{ip},\mathrm{cf}\}$; a channel \\
$\mathcal P$, $p$, $r_p$ & extraction-pass set $\{\mathrm{pref},\mathrm{ctx},\mathrm{reas},\mathrm{cf}\}$; a pass; raw output \\
$E=\{E_k\}_{k\in\mathcal K}$ & experience memory bank (Eq.~\ref{eq:bank}) \\
$E_k^{u}\subset E_k$ & channel-$k$ entries for user $u$ after metadata filter \\
$e=(\tau,\mathbf{v}_e,\mu)$ & memory entry: text, embedding $\mathbf{v}_e=\mathrm{Enc}(\tau)$, metadata \\
$\mathrm{Adapt}$, $\mathrm{route}_k$ & distill routed fields into $e_k$; select fields for channel $k$ \\
$\mathrm{snapshot}(E)$ & truncated text rendering of the bank for $\mathrm{LLM}_C$ \\
$B=\{(\tau,k,u)\}$ & insight batch (teacher or arbiter) consumed by $\mathrm{LLM}_C$ \\
$o$, $\mathrm{Apply}$ & edit ops $\{\textsc{Add},\textsc{Delete},\textsc{Modify},\textsc{Keep}\}$; commit $o$ to $E$ \\
$m$, $R_k(u,C_u)$ & retrieval depth; top-$m$ entries from $E_k^{u}$ (Eq.~\ref{eq:retrieval}) \\
$\hat L_u$, $\mathrm{rank}$ & student ranked list; 1-based position of $i^+$ in $\hat L_u$ \\
$r(u)$ & post-prediction reward vector (Eq.~\ref{eq:reward}) \\
$K$, $n_r$, $T$ & \#debate agents; rounds/epoch; optimization epochs \\
$\mathcal{T}$, $g_j$, $\mathrm{LLM}_j$ & debate transcript; critique of agent $j$; $j$-th debate agent \\
$\tilde{e}$, $n_{\max}$, $U$ & arbiter-proposed entry; max commits/case; case batch \\
$\mathrm{LLM}_T$, $\mathrm{LLM}_S$, $\mathrm{LLM}_C$, $\mathrm{LLM}_A$ & teacher, frozen student, controller, arbiter \\
\bottomrule
\end{tabularx}
\end{table}

\subsection{Problem Formulation}
\label{sec:method-formulation}

We cast next-item recommendation as \emph{conditional generation}: given user $u$, history $H_u$, and a candidate set $C_u$, a language model with parameters $\theta$ assigns a score to each candidate $c\in C_u$ by the likelihood of emitting $c$ under a constructed prompt. The predicted item is
\begin{equation}
\hat{i} = \operatorname*{arg\,max}_{c\,\in\,C_u} P_\theta\big(c \mid \mathrm{context}(u)\big), \label{eq:task}
\end{equation}
and is evaluated against the held-out positive $i^+$ using HR@$k$, NDCG@$k$, and MRR (Section~\ref{sec:setup-protocol}). Methods differ in what enters $\mathrm{context}(u)$; we make this explicit below.

\subsection{Preliminaries: Prior Paradigms as Conditional Generation}
\label{sec:method-prelim}

\textbf{Zero-/few-shot prompting}~\citep{brown2020gpt3} conditions the LLM on a natural-language prompt $x_u$ and $K$ in-context demonstrations $D_K=\{(x_j,y_j)\}_{j=1}^K$, with no parameter update:
\begin{equation}
P_{\mathrm{FS}}(c\mid u) = P_\theta\big(c \mid x_u, D_K\big), \quad x_u=\mathrm{prompt}(H_u, M(C_u)). \label{eq:fewshot}
\end{equation}
$K{=}0$ recovers zero-shot. Every token of $x_u$ and $D_K$ is re-encoded by $\theta$ at every request.

\textbf{RAG}~\citep{lewis2020rag} retrieves a top-$k$ evidence set $Z$ and conditions generation on it:
\begin{equation}
P_{\mathrm{RAG}}(c\mid u) = P_\theta\big(c \mid x_u, Z\big), \quad Z=\mathrm{Top}\text{-}k(x_u). \label{eq:rag}
\end{equation}
In our RAG baseline, $Z$ ranges over \emph{raw} interaction/review records and is recomputed for every $(u,C_u)$.

\textbf{GraphRAG}~\citep{edge2024graphrag} builds a corpus graph, partitions it into communities $\{S_l\}$ with LLM-written summaries $\sigma(S_l)$, and answers by a map--reduce over community answers $a_l$:
\begin{equation}
P_{\mathrm{GRAG}}(c\mid u) = \mathrm{LLM}\big(x_u, c, \{a_l\}_{l=1}^{L}\big), \quad a_l = \mathrm{LLM}\big(x_u, \sigma(S_l)\big). \label{eq:graphrag}
\end{equation}
$\{\sigma(S_l)\}$ is static once built and is not typed, per-user, or editable at the granularity of a single fact.

Equations~\eqref{eq:fewshot}--\eqref{eq:graphrag} share a property that motivates our design: the conditioning set ($D_K$, $Z$, or $\{\sigma(S_l)\}$) is either rebuilt/rescored at every request, or, once built, exposes no operator for targeted, entry-level edits. Section~\ref{sec:method-overview} replaces this with a persistent structure $E$ built off the inference critical path and equipped with an explicit edit operator (Section~\ref{sec:method-memory}).

\subsection{Overview}
\label{sec:method-overview}

\begin{figure*}[pos=t]
\centering
\includegraphics[width=\textwidth]{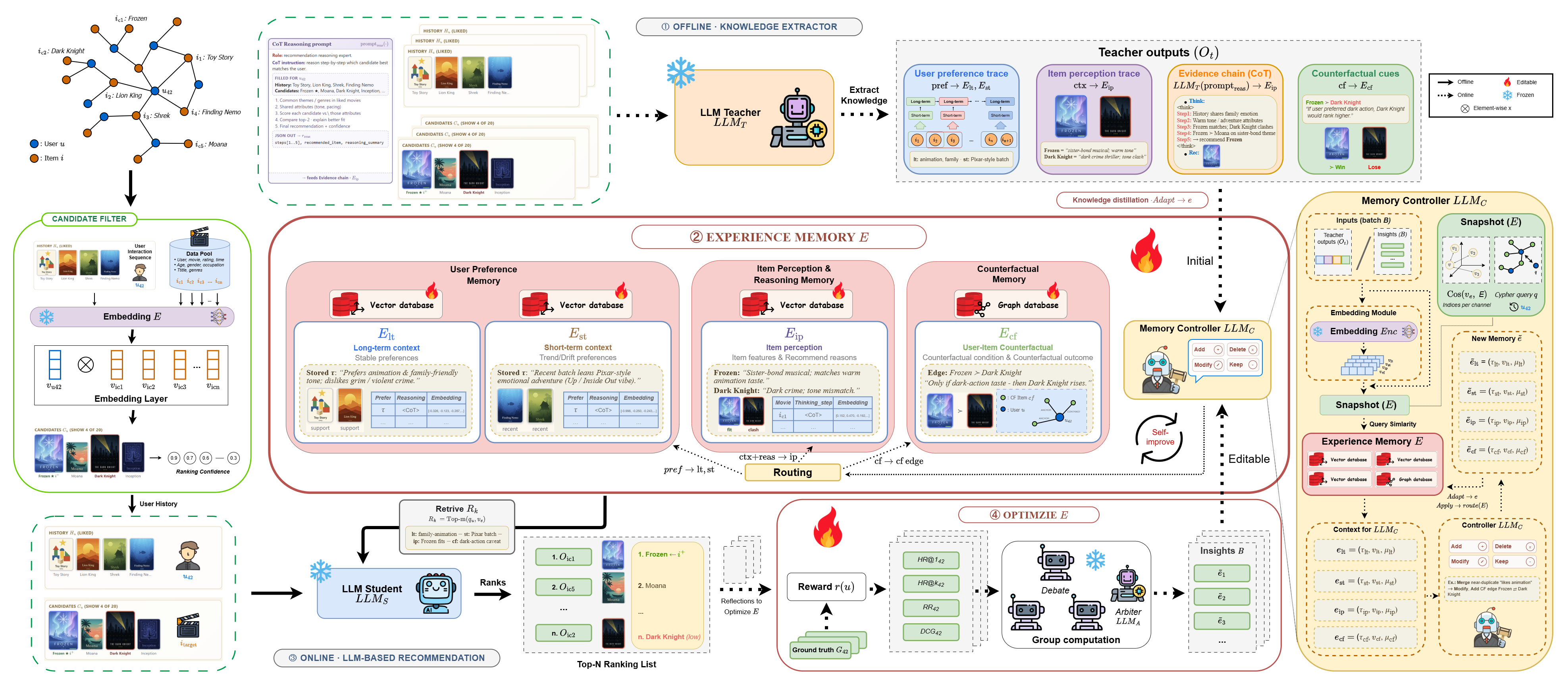}
\caption{Overall architecture of rEDMRec on a concrete case (target \textit{Frozen}, hard negative \textit{The Dark Knight}), aligned with Sections~\ref{sec:method-adapter}--\ref{sec:method-optimize}. \circled{1}~LLM Teacher as Knowledge Extractor: frozen $\mathrm{LLM}_T$ emits preference / perception / evidence / counterfactual traces; Knowledge Distillation ($\mathrm{Adapt}$) writes four-channel entries that the Memory Controller $\mathrm{LLM}_C$ commits via Add/Delete/Modify/Keep. \circled{2}~Editable Experience Memory Bank $E=\{E_{\mathrm{lt}},E_{\mathrm{st}},E_{\mathrm{ip}},E_{\mathrm{cf}}\}$. Candidate Filter forms $C_u$ with shared $\mathrm{Enc}$. \circled{3}~LLM Student as Ranker: frozen $\mathrm{LLM}_S$ retrieves top-$m$ entries per channel and ranks without calling the teacher. \circled{4}~Experience Memory Optimization: Reward Models $r(u)$ and Debate and Arbiter ($\mathrm{LLM}_A$) revise $E$ while $\mathrm{LLM}_S$ stays fixed.}
\label{fig:architecture}
\end{figure*}

rEDMRec factorizes ranking into an offline construction of a typed experience memory $E$ and an online, teacher-free generative lookup, as illustrated in Figure~\ref{fig:architecture}:
\begin{equation}
\begin{aligned}
P_S(c\mid u)
  &= P_S\Big(c \,\Big|\, x_u,\; {\textstyle\bigcup_{k\in\mathcal K}} R_k(u,C_u)\Big), \\
E &\leftarrow \mathrm{Apply}(E,\, o),
\end{aligned}
\label{eq:overview}
\end{equation}
where $P_S$ denotes generation under the frozen student $\mathrm{LLM}_S$, $R_k$ (Eq.~\ref{eq:retrieval}) is a deterministic top-$m$ dense lookup (not a per-request marginalized set as in Eq.~\eqref{eq:rag}), and $\mathrm{Apply}$ commits edit ops $o$ produced offline by teacher extraction (Section~\ref{sec:method-teacher}) and distillation (Section~\ref{sec:method-memory}), or online by debate-based optimization after each student prediction (Section~\ref{sec:method-optimize}). This is the central contrast with Eqs.~\eqref{eq:fewshot}--\eqref{eq:graphrag}: reasoning cost is paid when writing $E$, the frozen student only retrieves, and memory edits -- not weight updates -- amortize future ranking.

Concretely, Figure~\ref{fig:architecture} decomposes the pipeline into the Method subsections that follow. \circled{1}~LLM Teacher as Knowledge Extractor (Section~\ref{sec:method-teacher}): given $(u,H_u,i,M)$, $\mathrm{LLM}_T$ runs four extraction passes and Knowledge Distillation $\mathrm{Adapt}$ (Section~\ref{sec:method-distill}) normalizes each routed signal into a channel-indexed entry $e=(\tau,\mathbf{v}_e,\mu)$, which the Memory Controller $\mathrm{LLM}_C$ commits with Add/Delete/Modify/Keep. \circled{2}~Editable Experience Memory Bank (Section~\ref{sec:method-memory}): $E=\{E_{\mathrm{lt}},E_{\mathrm{st}},E_{\mathrm{ip}},E_{\mathrm{cf}}\}$ stores stable taste, session context, candidate-grounded perception, and counterfactual hard-negative edges as independently retrievable typed snippets. Before online ranking, Candidate Filter (Section~\ref{sec:method-adapter}) forms $C_u$ and shares $\mathrm{Enc}$ with memory indexing. \circled{3}~LLM Student as Ranker (Section~\ref{sec:method-student}): frozen $\mathrm{LLM}_S$ encodes the current query, retrieves the top-$m$ entries per channel, and produces $\hat L_u$ without invoking $\mathrm{LLM}_T$ or updating student weights. \circled{4}~Experience Memory Optimization (Section~\ref{sec:method-optimize}): Reward Models $r(u)$ score $\hat L_u$ against $i^+$, then Debate and Arbiter $\mathrm{LLM}_A$ propose revisions that $\mathrm{Adapt}$ and $\mathrm{Apply}$ write back into $E$, so future retrievals improve while $\mathrm{LLM}_S$ stays fixed.

\subsection{Candidate Filter}
\label{sec:method-adapter}

\emph{Motivation:} a generative student cannot score the full catalog $\mathcal I$ at each request, and retrieval plus memory indexing must share one dense text space rather than operating over raw strings. \emph{Design:} candidate sets $C_u$ are formed by a lightweight recency/popularity/retrieval-score filter over $\mathcal I$ (not a learned projector). The same encoder maps any text span to a $d$-dimensional retrieval vector,
\begin{equation}
\mathbf{v} = \mathrm{Enc}(\mathrm{text}), \qquad
\mathrm{sim}(\mathbf{v},\mathbf{v}') = \frac{\mathbf{v}^{\top}\mathbf{v}'}{\|\mathbf{v}\|\,\|\mathbf{v}'\|}, \label{eq:adapter}
\end{equation}
with $\mathrm{Enc}$ a sentence-transformer~\citep{reimers2019sbert}. $\mathbf{v}$ is reused for memory indexing (Eq.~\ref{eq:entry}) and retrieval (Eq.~\ref{eq:retrieval}). \emph{Advantage:} $C_u$ is formed without a learned projector, and one shared $\mathrm{Enc}$ lets $E$ be queried directly with candidate or user vectors, with no separate alignment step.

\subsection{LLM Teacher as Knowledge Extractor}
\label{sec:method-teacher}

\emph{Motivation:} conflating ``understand the user'' (stable) with ``score this candidate list'' (per-request) forces the former to be redone every request; rEDMRec queries the teacher only for the former, and only to \emph{extract}, never to rank. \emph{Design:} following ReasoningRec~\citep{bismay2025reasoningrec} and R2Rec~\citep{r2rec}, which show that structured preference profiles, item-level perceptions, and explanatory rationales improve recommendation when distilled from a strong LLM, teacher $\mathrm{LLM}_T$ runs four extraction passes $p\in\mathcal P$, each a prompted generation,
\begin{equation}
\begin{aligned}
r_p
  &= \mathrm{LLM}_T\big(\mathrm{prompt}_p(u, i, H_u, M)\big), \\
  &\qquad p\in\mathcal P.
\end{aligned}
\label{eq:teacher}
\end{equation}
whose output fields are later routed to one or more memory channels $k\in\mathcal K$ by $\mathrm{Adapt}$ (Table~\ref{tbl:signals}, Section~\ref{sec:method-distill}); the routing is not one-to-one, since two passes jointly populate the item-perception channel. Prompt templates and example one-line outputs for all four channels are given in Appendix~\ref{app:prompts}; full Beauty/Steam prediction traces (history, candidates, $R_k$, ranked list) appear in Appendix~\ref{app:examples}. Unlike ReasoningRec/R2Rec, which consume these traces as training targets or one-shot prompt features, we never update student parameters from $r_p$: each field is committed into the editable bank $E$ of Section~\ref{sec:method-memory}.

\begin{table}[pos=t]
\caption{Teacher extraction passes $p\in\mathcal P$: output fields and routed channel(s) $k$ (Section~\ref{sec:method-distill}).}
\label{tbl:signals}
\footnotesize
\centering
\begin{tabularx}{\columnwidth}{@{}lYl@{}}
\toprule
$p$ & Key output fields & Routed to $k$ \\
\midrule
$\mathrm{pref}$ & \texttt{long\_term\_preferences}, \texttt{dislikes}, \texttt{short\_term\_preferences} & $\mathrm{lt}$, $\mathrm{st}$ \\
$\mathrm{ctx}$ & \texttt{user\_history\_perception}, \texttt{candidate\_perception} & $\mathrm{ip}$ \\
$\mathrm{reas}$ & \texttt{steps[1..5]}, \texttt{reasoning\_summary} & $\mathrm{ip}$ \\
$\mathrm{cf}$ & \texttt{counterfactual\_condition/outcome}, \texttt{rationale} & $\mathrm{cf}$ \\
\bottomrule
\end{tabularx}
\end{table}

\textbf{Preference extraction ($\mathrm{pref}$).} Simulates a recurrent, batch-by-batch update over $H_u$ (oldest$\to$newest): for each history batch it re-estimates a long-term preference state and the current-batch short-term interest, plus an explicit dislike list; the final long-term/dislike fields route to channel $\mathrm{lt}$ and the short-term field routes to channel $\mathrm{st}$.

\textbf{Context extraction ($\mathrm{ctx}$).} For every history item and candidate, produces three layers -- an objective factual description, a first-person ``as this user'' comment, and (candidates only) key phrases; both the history-perception and candidate-perception fields route to the item-perception channel $\mathrm{ip}$.

\textbf{Reasoning extraction ($\mathrm{reas}$).} Runs a five-step chain-of-thought that identifies shared themes across liked items, scores each candidate against them, contrasts the top two candidates, and states a final recommendation; its summary field also routes to $\mathrm{ip}$, complementing $\mathrm{ctx}$ with an explicit comparative rationale rather than a per-item description.

\textbf{Counterfactual extraction ($\mathrm{cf}$).} Given an anchor (chosen) item and a contrast (hard-negative) item, produces why-preferred / why-rejected rationales and a hypothetical condition under which the contrast item would outrank the anchor; routes entirely to channel $\mathrm{cf}$ as a graph edge (Section~\ref{sec:method-channels}).

An optional single-pass multi-agent refinement critiques the $r_{\mathrm{cf}}$ output with three fixed personas before distillation, merging by ``last full critique wins'' with no arbiter; this is a lighter-weight precursor to the arbiter-based optimization of Section~\ref{sec:method-optimize}, which instead revises already-committed entries across all four channels using downstream reward signals. \emph{Advantage:} separating extraction into four typed passes, rather than one undifferentiated call, lets Section~\ref{sec:method-distill} edit a single routed claim without touching the other channels of the same interaction.

\subsection{Editable Experience Memory Bank}
\label{sec:method-memory}

\emph{Motivation:} Eq.~\eqref{eq:teacher} yields free-form JSON that cannot be reused across sessions or revised entry-by-entry once committed. \emph{Design:} adapting the training-free experiential-knowledge library of Training-Free GRPO~\citep{chen2025trainingfreegrpo} to recommendation, rEDMRec materializes teacher output into a persistent, typed non-parametric memory
\begin{equation}
E = \{E_k\}_{k\in\mathcal K}, \quad E_k = \{e\mid e.\mathrm{memory\_type}=k\}, \label{eq:bank}
\end{equation}
where each entry $e=(\tau,\mathbf{v}_e,\mu)$ carries distilled text $\tau$, embedding $\mathbf{v}_e=\mathrm{Enc}(\tau)$, and channel-specific metadata $\mu$. Population follows a two-stage commit path -- $\mathrm{Adapt}$ (field routing and schema normalization) and $\mathrm{Apply}$ via controller $\mathrm{LLM}_C$ (batch edit ops) -- detailed in Section~\ref{sec:method-distill}; channel semantics and retrieval are in Section~\ref{sec:method-channels}.

\subsubsection{Knowledge Distillation: Adapter and Memory Controller}
\label{sec:method-distill}

\textbf{Distillation adapter $\mathrm{Adapt}$.} $\mathrm{Adapt}$ maps routed teacher fields (Table~\ref{tbl:signals}) into a typed entry and dispatches it to the correct physical store:
\begin{equation}
\begin{aligned}
e_k
  &= \mathrm{Adapt}\big(\mathrm{route}_k(\{r_p\}_{p\in\mathcal P})\big) \\
  &= \big(\tau_k,\; \mathrm{Enc}(\tau_k),\; \mu_k\big),
     \qquad k\in\mathcal K.
\end{aligned}
\label{eq:entry}
\end{equation}
Here $\mathrm{route}_k(\cdot)$ selects and concatenates the field(s) assigned to channel $k$ (many-to-one: $\mathrm{pref}\!\to\!\{\mathrm{lt},\mathrm{st}\}$; $\mathrm{ctx},\mathrm{reas}\!\to\!\mathrm{ip}$; $\mathrm{cf}\!\to\!\mathrm{cf}$). $\mathrm{Adapt}$ then (i) normalizes $\tau_k$ into a channel schema ($\mu_k$ holds typed fields such as \texttt{long\_term\_preferences}, \texttt{dislikes}, \texttt{steps}, \texttt{anchor\_item}), (ii) encodes $\tau_k$ with the shared $\mathrm{Enc}$ of Eq.~\eqref{eq:adapter}, and (iii) writes to $E_{\mathrm{lt}}$, $E_{\mathrm{st}}$, or $E_{\mathrm{ip}}$ (Vector database) or, for $k=\mathrm{cf}$, to $E_{\mathrm{cf}}$ as a Graph database  edge pair $(u)\!\xrightarrow{\textsc{Anchor}}\!(i^+)\wedge(u)\!\xrightarrow{\textsc{Contrast}}\!(i^-)$ plus a rationale embedding (Listing~\ref{lst:counterfactual}). Initial extraction (\texttt{distill\_to\_memory}) and debate optimization (Section~\ref{sec:method-optimize}) both call the same $\mathrm{Adapt}$.

\textbf{Memory controller $\mathrm{LLM}_C$.} Raw $\mathrm{Adapt}$ output is not appended blindly: controller $\mathrm{LLM}_C$ inspects a text snapshot of the bank against a batch of new insights $B=\{(\tau,\,k,\,u)\}$ and emits structured edit ops $o$, which $\mathrm{Apply}$ commits in order:
\begin{equation}
\begin{aligned}
o &= \mathrm{LLM}_C\big(\mathrm{snapshot}(E),\, B\big), \\
o_i &\in \{\textsc{Add}, \textsc{Delete}, \textsc{Modify}, \textsc{Keep}\}, \\
E &\leftarrow \mathrm{Apply}(E,\, o).
\end{aligned}
\label{eq:controller}
\end{equation}
$\mathrm{Apply}$ implements each op on the underlying stores: \textsc{Add} calls $\mathrm{Adapt}$ then appends; \textsc{Delete} removes by \texttt{entry\_id}; \textsc{Modify} re-encodes revised $\tau$ and replaces metadata in-place; \textsc{Keep} is a no-op. These four operators mirror the experience-library update rule in Training-Free GRPO~\citep{chen2025trainingfreegrpo}, specialized here to typed recommendation channels rather than general agent rollouts. The controller prompt enforces a maximum library size, deduplication (merge over add), and $\le$60-word actionable entries. \emph{Advantage:} Eqs.~\eqref{eq:entry}--\eqref{eq:controller} decouple \emph{what} to remember (teacher or arbiter) from \emph{how} to commit it; refinement cost scales with $|o|$, not $|E|$.

\subsubsection{Four Experience-Memory Channels}
\label{sec:method-channels}

We factor experience into four typed channels because prior LLM and sequential recommenders show that long-horizon taste~\citep{shi2024large,wang2025leveraging}, recent sequential context~\citep{zheng2024llmtrsr,zhang2024lsidn}, item-level semantics~\citep{ren2024rlmrec,zhang2026token2item}, and hard-negative contrast~\citep{song2026llmhni,li2026ilrec} each contribute distinct ranking signal -- and because storing them separately lets $R_k$ retrieve only the evidence class needed for a decision. Table~\ref{tbl:channels} summarizes the four channels. Each $E_k$ is independently indexed, independently retrievable, and independently editable -- the ablation in Section~\ref{sec:results-h2} drops individual $k$ at inference without retraining $\mathrm{LLM}_S$.

\begin{table}[pos=t]
\caption{Four experience-memory channels: role, teacher source, storage, and retrieval filter.}
\label{tbl:channels}
\footnotesize
\centering
\begin{tabularx}{\columnwidth}{@{}lYlY@{}}
\toprule
$k$ & Semantic role & Source $p$ & Storage / filter \\
\midrule
$\mathrm{lt}$ & Stable cross-session taste & $\mathrm{pref}$ & vector DB; \texttt{user\_id}$=u$ \\
$\mathrm{st}$ & Session-level trend / drift & $\mathrm{pref}$ & vector DB + timestamp; \texttt{user\_id}$=u$ \\
$\mathrm{ip}$ & Item impression vs.\ user & $\mathrm{ctx}$, $\mathrm{reas}$ & vector DB; \texttt{user\_id}$=u$, \texttt{item}$=c$ \\
$\mathrm{cf}$ & Contrastive ``if\ldots'' edge & $\mathrm{cf}$ & graph DB + vector DB; \texttt{user\_id}$=u$ \\
\bottomrule
\end{tabularx}
\end{table}

\textbf{Long-term preference ($\mathrm{lt}$).} Motivated by evidence that long-horizon preference modeling improves recommendation beyond next-click prediction~\citep{shi2024large} and that external memory retrieval is needed when the LLM context window alone drops long-term history~\citep{wang2025leveraging}, this channel stores durable taste statements distilled from \texttt{long\_term\_preferences} and \texttt{dislikes} ($\mathrm{pref}$ pass), optionally with supporting \texttt{reasoning}. Entries are user-global: retrieval over $E_{\mathrm{lt}}^{u}=\{e\in E_{\mathrm{lt}}\mid\mu.\texttt{user\_id}=u\}$ answers ``what does this user generally like/dislike?'' without binding to a specific candidate.

\textbf{Short-term context ($\mathrm{st}$).} Motivated by sequential recommenders that show recent interactions dominate next-item prediction~\citep{zheng2024llmtrsr} and that jointly modeling long- and short-term interests outperforms either alone~\citep{zhang2024lsidn}, this channel captures transient interest from \texttt{short\_term\_preferences} ($\mathrm{pref}$ pass) plus optional \texttt{context\_reasoning}. Each entry carries a timestamp in $\mu$, enabling the bank to represent taste \emph{drift} within a session or across recent batches while $\mathrm{lt}$ remains stable. Retrieval filters on $\texttt{user\_id}$ only.

\textbf{Item perception ($\mathrm{ip}$).} Motivated by work showing that item-text representation quality drives LLM recommendation~\citep{ren2024rlmrec} and that token-centric attention under-models item-level collaborative relations~\citep{zhang2026token2item}, this is the most granular channel: (i) per-candidate impressions from \texttt{candidate\_perception} ($\mathrm{ctx}$), (ii) per-history-item descriptions from \texttt{user\_history\_perception} ($\mathrm{ctx}$), and (iii) comparative reasoning from the five-step chain and \texttt{reasoning\_summary} ($\mathrm{reas}$). Candidate-specific entries additionally index $\texttt{target\_movie\_id}$ in $\mu$, so retrieval for $(u,c)$ returns impressions scoped to $c$ rather than unrelated items.

\textbf{Counterfactual ($\mathrm{cf}$).} Motivated by recent LLM recommenders that separate hard negatives from noisy negatives~\citep{song2026llmhni} and that use self-hard negatives to sharpen preference learning~\citep{li2026ilrec}, this channel stores hard-negative contrast rationales as typed graph edges: for anchor item $i^+$ and contrast $i^-$, $\mu$ records \texttt{why\_anchor\_preferred}, \texttt{why\_contrast\_rejected}, \texttt{counterfactual\_condition}, and \texttt{counterfactual\_outcome}. Dense retrieval runs over \texttt{rationale\_text} embeddings; a graph-augmented pass appends any user-specific edges not surfaced by vector search. Listing~\ref{lst:counterfactual} shows the instantiated schema.

\textbf{Retrieval.} At inference, the student queries each active channel with a shared query embedding $\mathbf{q}_{u}=\mathrm{Enc}(\mathrm{prompt}(u,C_u))$ and returns the top-$m$ entries by dense similarity:
\begin{equation}
R_k(u,C_u)
  = \operatorname*{Top\text{-}m}_{e\,\in\,E_k^{u}}
    \mathrm{sim}\big(\mathbf{q}_{u},\, \mathbf{v}_e\big), \label{eq:retrieval}
\end{equation}
where $E_k^{u}\subseteq E_k$ applies the filters in Table~\ref{tbl:channels} (candidate scope $c$ is applied only for $\mathrm{ip}$), and $\mathrm{sim}$ is Eq.~\eqref{eq:adapter}. Default $m{=}5$ per channel. \emph{Advantage:} typed channels let $R_k$ return only the evidence class needed for a ranking decision -- stable taste ($\mathrm{lt}$), recent drift ($\mathrm{st}$), candidate fit ($\mathrm{ip}$), or boundary conditions ($\mathrm{cf}$) -- rather than one undifferentiated reasoning blob.

\begin{figure}[pos=t]
\footnotesize
\begin{verbatim}
{
 "user_id": "...",
 "anchor_preference":
   "prefers animation,
    family-friendly tone",
 "target_item": "Frozen",
 "contrast_item":
   "The Dark Knight",
 "counterfactual_condition":
   "if user prefers dark,
    action-heavy stories",
 "counterfactual_outcome":
   "The Dark Knight would
    rank higher",
 "rationale_text": "...",
 "embedding": [],
 "timestamp": "..."
}
\end{verbatim}
\caption{Distilled $e_{\mathrm{cf}}$: the graph-edge instantiation of Eq.~\eqref{eq:entry}.}
\label{lst:counterfactual}
\end{figure}

\subsection{LLM Student as Ranker}
\label{sec:method-student}

\emph{Motivation:} if ranking gains required adapting student parameters, it would be unclear whether accuracy came from the editable experience memory or from parameter adaptation; moreover, a per-backbone adapted student would need re-adaptation for every backbone, undermining the claim that $E$ helps \emph{any} teacher--student pair. \emph{Design:} student $\mathrm{LLM}_S$ is a frozen pretrained LLM (3B--20B in our study). At inference it ranks solely by prompting with retrieved experience -- never a fresh teacher call and never a weight update -- following Eq.~\eqref{eq:overview}:
\begin{equation}
\begin{aligned}
\hat{i}
  &= \operatorname*{arg\,max}_{c\,\in\,C_u} P_S(c\mid u) \\
  &= \operatorname*{arg\,max}_{c\,\in\,C_u}
     P_S\Big(c \,\Big|\, x_u,\; {\textstyle\bigcup_{k\in\mathcal K}} R_k(u,C_u)\Big).
\end{aligned}
\label{eq:student-rank}
\end{equation}
Here $x_u=\mathrm{prompt}(u,C_u)$ and $\bigcup_k R_k$ are concatenated into the student context; $\mathrm{LLM}_S$ is held fixed across bank updates. Unlike Eq.~\eqref{eq:rag}, retrieval is deterministic, and no student objective is optimized. \emph{Advantage:} gains over zero-/few-shot and RAG -- and over GraphRAG on most backbones (Section~\ref{sec:results-h1}) -- are attributable to the content and editability of $E$, not to student adaptation -- the same frozen $\mathrm{LLM}_S$ improves when $E$ improves (Section~\ref{sec:results-h4}), and the protocol transfers across teacher--student pairs without per-student training (Section~\ref{sec:results-h3}).

\subsection{Experience Memory Optimization}
\label{sec:method-optimize}

\emph{Motivation:} a bank populated once from teacher extraction can accumulate generic or conflicting entries; after the student ranks a case, the mismatch between the predicted list and the ground-truth target is a direct signal that $E$ should be revised. \emph{Design:} mirroring Training-Free GRPO~\citep{chen2025trainingfreegrpo}'s loop of rollout $\to$ reward $\to$ semantic advantage $\to$ library edit -- but specialized to recommendation ranking and enriched with $K$-agent debate -- we update $E$ \emph{after each student prediction}: the student emits a ranked list under Eq.~\eqref{eq:student-rank}, reward models score that list against $i^+$, $K$ debating agents critique the case conditioned on those rewards, an arbiter synthesizes revised experience entries, and the same distillation controller commits them. This closes a loop $\mathrm{predict}\!\to\!\mathrm{reward}\!\to\!\mathrm{debate}\!\to\!\mathrm{edit}\!\to\!E$ without changing $\mathrm{LLM}_S$.

\subsubsection{Reward Models}
\label{sec:method-reward}

\emph{Motivation:} a single scalar (e.g., only Hit@1) is too coarse for debate agents to diagnose \emph{why} a ranking failed -- missing the top item, burying it just outside the top-$k$, or placing it deep in the list require different memory edits. \emph{Design:} given the student's ordered list $\hat L_u=(c_{(1)},\ldots,c_{(L)})$ and ground-truth target $i^+$, let $\mathrm{rank}\in\{1,\ldots,L\}\cup\{\varnothing\}$ be the 1-based position of $i^+$ in $\hat L_u$ ($\varnothing$ if absent). We pack four complementary IR-style rewards in $[0,1]$ into the debate/arbiter prompt:
\begin{equation}
\begin{aligned}
r(u)
  &= \big(
      \mathbf{1}[\mathrm{rank}{=}1],\;
      \mathbf{1}[\mathrm{rank}\le k], \\
  &\qquad
      \min(1,1/\mathrm{rank}),\;
      1/\log_2(\mathrm{rank}+1)
     \big)
     \in[0,1]^4,
\end{aligned}
\label{eq:reward}
\end{equation}
with all components zero if $\mathrm{rank}=\varnothing$ (default $k{=}10$). \textbf{Design rationale.} The four components are HR@1, HR@$k$, reciprocal rank, and DCG-style position discount: they flag strict top-rank failure, near-misses, and graded credit when $i^+$ appears but not first. Importantly, $r(u)$ is \emph{not} a training loss on $\mathrm{LLM}_S$ and is \emph{not} used to $\arg\max$ among agent proposals; it conditions the natural-language critique so that edits target the observed ranking failure mode. \emph{Advantage:} the same four signals transfer across datasets and students because they depend only on $(\hat L_u,i^+)$, keeping the optimization loop student-agnostic.

\subsubsection{Debate and Arbiter after Each Prediction}
\label{sec:method-debate}

\textbf{Design lineage.} The procedure combines the training-free experience-library update of Training-Free GRPO~\citep{chen2025trainingfreegrpo} with iterative self-critique~\citep{madaan2023selfrefine} and multi-persona debate~\citep{du2023multiagentdebate}, but applies them at memory-bank granularity and triggers them from the student's post-prediction reward vector (Section~\ref{sec:method-reward}).

\textbf{Group computation.} Over $n_r$ rounds, $K$ fixed-persona agents append free-form critiques to a shared transcript $\mathcal{T}$:
\begin{equation}
\begin{aligned}
g_j &= \mathrm{LLM}_j\big(\mathrm{persona}_j,\, \mathcal{T},\, u,\, \hat L_u,\, r(u)\big), \\
\mathcal{T} &\leftarrow \mathcal{T} \,\Vert\, g_j.
\end{aligned}
\label{eq:debate}
\end{equation}
No entry is committed during Eq.~\eqref{eq:debate}. After $n_rK$ turns, a single arbiter call synthesizes the experience set to commit:
\begin{equation}
\begin{aligned}
\{\tilde{e}_1,\ldots,\tilde{e}_n\}
  &= \mathrm{LLM}_A\big(\mathcal{T},\, u,\, \hat L_u,\, r(u)\big), \\
  &\qquad n\le n_{\max}.
\end{aligned}
\label{eq:arbiter}
\end{equation}
Eq.~\eqref{eq:arbiter} is an LLM synthesis, not a programmatic vote or an $\arg\max_j$ over $\{g_j\}$ by reward.

\textbf{Optimizing.} Algorithm~\ref{alg:debate} runs, for each case: student ranking (Eq.~\ref{eq:student-rank}) $\to$ reward models (Eq.~\ref{eq:reward}) $\to$ debate/arbiter (Eqs.~\ref{eq:debate}--\ref{eq:arbiter}) $\to$ commit via $\mathrm{Adapt}$ and $\mathrm{Apply}$. Section~\ref{sec:results-h4} tracks bank-quality signals and downstream HR@1/MRR over $T$ such epochs.

\textbf{Reuse of Knowledge Distillation.} Each synthesized $\tilde{e}$ is committed by the \emph{same} operators as regular extraction (Section~\ref{sec:method-distill}): $e=\mathrm{Adapt}(\tilde{e})$ (Eq.~\ref{eq:entry}) and $E \leftarrow \mathrm{Apply}(E, o)$ (Eq.~\ref{eq:controller}). Optimization does not update $\mathrm{LLM}_S$; it only revises $E$.

\begin{algorithm}[t]
\footnotesize
\caption{Experience Memory Optimization after student prediction (one epoch)}
\label{alg:debate}
\begin{algorithmic}[1]
\Require case batch $U$, frozen student $\mathrm{LLM}_S$, agents $\{\mathrm{LLM}_1,\ldots,\mathrm{LLM}_K\}$, arbiter $\mathrm{LLM}_A$, rounds $n_r$
\For{$u \in U$}
  \State Rank $\hat L_u$ with frozen $\mathrm{LLM}_S$ via Eq.~\eqref{eq:student-rank}
  \State $r(u) \gets$ reward models on $(\hat L_u, i^+)$ \Comment{Eq.~\eqref{eq:reward}}
  \State $\mathcal{T} \gets \varnothing$
  \For{round $=1,\ldots,n_r$; agent $j=1,\ldots,K$}
    \State $\mathcal{T} \gets \mathcal{T} \,\Vert\, \mathrm{LLM}_j(\mathrm{persona}_j,\mathcal{T},u,\hat L_u,r(u))$ \Comment{Eq.~\eqref{eq:debate}}
  \EndFor
  \State $\{\tilde{e}_1,\ldots,\tilde{e}_n\} \gets \mathrm{LLM}_A(\mathcal{T}, u, \hat L_u, r(u))$ \Comment{Eq.~\eqref{eq:arbiter}}
  \For{$\tilde{e} \in \{\tilde{e}_1,\ldots,\tilde{e}_n\}$}
    \State $e \gets \mathrm{Adapt}(\tilde{e})$; \; $E \gets \mathrm{Apply}(E, o)$ \Comment{Eqs.~\eqref{eq:entry},\eqref{eq:controller}}
  \EndFor
\EndFor
\end{algorithmic}
\end{algorithm}

\section{Experimental Setup}
\label{sec:setup}

\subsection{Datasets}
\label{sec:setup-datasets}

We evaluate on three recommendation datasets that span the explicit-implicit feedback spectrum. \textbf{ML-1M}~\citep{harper2015movielens} provides 1--5 star explicit ratings and is a standard benchmark for LLM-based recommendation. \textbf{Amazon Beauty}~\citep{ni2019amazon} provides explicit ratings and review text but is substantially sparser per user, which stresses the memory bank's ability to compensate for weak collaborative signal. \textbf{Steam}~\citep{kang2018steam} is an \emph{implicit-feedback} dataset: it logs which games a user played rather than an explicit rating, so every logged interaction is treated as an equally-weighted positive signal (rating fixed to $1.0$, positive threshold $0.0$) instead of being thresholded from a graded rating scale. For all three datasets we apply a $k$-core filter ($k{=}20$ on ML-1M; $k{=}5$ on Beauty and Steam), build a chronological train/validation/test split, and construct 20-candidate ranking samples (1 positive, 19 sampled negatives). Dataset statistics after filtering are reported in Appendix~\ref{app:dataset-stats} (Table~\ref{tbl:app-dataset-stats}).

\subsection{Baselines}
\label{sec:setup-baselines}

We compare against four prompting/retrieval baselines and rEDMRec, formalized in Eqs.~\eqref{eq:fewshot}--\eqref{eq:graphrag} (Section~\ref{sec:method-prelim}): \textbf{Zero-shot} and \textbf{Few-shot}~\citep{brown2020gpt3} ($K{=}0$ and $K{>}0$ in Eq.~\ref{eq:fewshot}); \textbf{RAG}~\citep{lewis2020rag} (Eq.~\ref{eq:rag}), which retrieves raw historical interactions or reviews rather than distilled reasoning; and \textbf{GraphRAG}~\citep{edge2024graphrag} (Eq.~\ref{eq:graphrag}), which retrieves from a co-occurrence/knowledge graph built over items.

\subsection{Models}
\label{sec:setup-models}

Unless noted otherwise, the teacher is \texttt{gpt-5.4-mini}. We evaluate ten open student backbones spanning 3B--20B parameters: Qwen2.5 3B, Llama 3.1 8B~\citep{touvron2023llama}, Gemma-4-12B~\citep{team2024gemma}, Minimax M2.5, Mixtral 8x7B~\citep{jiang2024mixtral}, Qwen3-14B~\citep{yang2024qwen2}, DeepSeek-R1-Distill-Qwen-14B~\citep{guo2025deepseekr1}, Phi-4, Llama 4 Scout, and GPT OSS 20B. Each student is used \emph{frozen} (Section~\ref{sec:method-student}), so that cross-backbone gains isolate the contribution of the editable experience memory. The teacher-distillation study (Section~\ref{sec:results-h3}) additionally fixes the student to \texttt{gpt-5-mini} (strong) or Qwen2.5 3B (small) while varying the teacher across seven backbones, to isolate the teacher's contribution from the student's.

\subsection{Metrics and Protocol}
\label{sec:setup-protocol}

We report Hit Rate at rank $k$ (HR@$k$, $k\in\{1,5,10\}$), Normalized Discounted Cumulative Gain (NDCG@$k$, $k\in\{5,10\}$), and Mean Reciprocal Rank (MRR), all computed over the fixed 20-candidate set per evaluation sample. For RQ1 tables we report Impv (\%), the relative HR@1 improvement of rEDMRec over the second-best baseline on the same student ($\mathrm{Impv}=(\mathrm{Ours}-\mathrm{SecondBest})/\mathrm{SecondBest}\times 100$), following RDRec~\citep{wang2024rdrec}, together with a McNemar $p$-value on HR@1 vs.\ that second-best method (approximate $2{\times}2$ contingency reconstructed from the table HR@1 rates at the full held-out sizes $n_{\mathrm{ML-1M}}{=}49893$, $n_{\mathrm{Beauty}}{=}1460$, $n_{\mathrm{Steam}}{=}1460$; $^*$ marks $p{<}0.05$ when rEDMRec is ahead). Every (model, method, dataset) cell is evaluated on the \emph{full held-out test split} (chronological train/validation/test; 20 candidates per sample with sampling seed $42$), not a pilot subsample. The cross-dataset comparison in Section~\ref{sec:results-h1} (Amazon Beauty, Steam), the bank-scale analysis in Section~\ref{sec:results-h2}, and the $k$-EPOCH and number-of-agents curves in Section~\ref{sec:results-h4} follow the same full-split evaluation protocol at each reported dataset, bank scale, and epoch/agent-count setting.

\section{Results}
\label{sec:results}

We organize results around four research questions: does distilling reasoning into memory improve ranking over prompting-only and retrieval-only baselines (RQ1, Section~\ref{sec:results-h1})? which memory channel drives that improvement (RQ2, Section~\ref{sec:results-h2})? does teacher quality causally affect bank quality and downstream gain (RQ3, Section~\ref{sec:results-h3})? and does debate-based memory optimization measurably improve bank quality and downstream ranking (RQ4, Section~\ref{sec:results-h4})? We close with a qualitative case study of how individual memory entries evolve over training (Section~\ref{sec:results-qualitative}).

\subsection{RQ1: Does rEDMRec Improve Ranking Across Students and Datasets?}
\label{sec:results-h1}

Table~\ref{tbl:main-results} reports HR@$k$, NDCG@$k$, and MRR for four representative students spanning our capacity range (Qwen2.5 3B, Llama 3.1 8B, Mixtral 8x7B, Qwen3-14B) on ML-1M; the full ten-model table is given in Appendix~\ref{app:table1full}. rEDMRec improves HR@1 over Zero-shot, Few-shot, and RAG for every student. Following RDRec~\citep{wang2024rdrec}, Table~\ref{tbl:main-results} reports Impv (\%) vs.\ the second-best baseline (typically GraphRAG): the largest relative gain is on the smallest student (Qwen2.5 3B, Impv $=13.3\%$ over GraphRAG), consistent with structured memory helping most when parametric capacity is limited. Llama~3.1~8B is the clearest failure case -- Impv is negative because GraphRAG remains ahead (Impv $=-11.1\%$) -- which we attribute to weak instruction-following rather than a deficiency of the memory itself (Section~\ref{sec:setup-models}; Section~\ref{sec:discussion}); GPT~OSS~20B similarly trails GraphRAG slightly (Impv $=-3.3\%$), so the claim relative to GraphRAG is \emph{most}, not \emph{all}, students.

\begin{table*}[width=\FullWidth,pos=t]
\caption{Main results on ML-1M for four representative students (full ten-model table in Appendix~\ref{app:table1full}). Impv (\%) is the relative HR@1 gain of rEDMRec over the \emph{second-best} baseline on the same student, $\mathrm{Impv}=(\mathrm{Ours}-\mathrm{SecondBest})/\mathrm{SecondBest}\times 100$, following RDRec~\citep{wang2024rdrec}. $p$ is the exact McNemar $p$-value for rEDMRec HR@1 vs.\ the second-best baseline (full held-out $n{=}49893$; approximate contingency from the table HR@1 rates). $^*$ marks $p{<}0.05$ with rEDMRec ahead. Best in \textbf{bold}, second-best \underline{underlined}; rEDMRec rows labeled in bold.}
\label{tbl:main-results}
\footnotesize
\centering
\setlength{\tabcolsep}{3.2pt}
\begin{tabular}{@{}ll cccccc cc@{}}
\toprule
Model & Method & HR@1$\uparrow$ & HR@5$\uparrow$ & HR@10$\uparrow$ & NDCG@5$\uparrow$ & NDCG@10$\uparrow$ & MRR$\uparrow$ & Impv (\%) & $p$ \\
\midrule
Qwen2.5 3B & Zero-shot & 0.12 & 0.28 & 0.38 & 0.20 & 0.23 & 0.18 & -- & -- \\*
  & Few-shot & 0.14 & 0.30 & 0.40 & 0.21 & 0.24 & \second{0.20} & -- & -- \\*
  & RAG & 0.13 & 0.29 & 0.39 & 0.20 & 0.23 & 0.19 & -- & -- \\*
  & GraphRAG & \second{0.15} & \second{0.31} & \second{0.41} & \second{0.22} & \second{0.25} & \second{0.20} & -- & -- \\*
  & \makecell[l]{\textbf{rEDMRec}\\\textbf{(ours)}} & \best{0.17$^*$} & \best{0.35} & \best{0.45} & \best{0.25} & \best{0.28} & \best{0.23} & \best{+13.3$^*$} & $<0.001^*$ \\
\cmidrule(lr){1-10}
 Llama 3.1 8B & Zero-shot & 0.07 & 0.15 & 0.23 & \second{0.12} & 0.14 & \second{0.12} & -- & -- \\*
  & Few-shot & \second{0.08} & \second{0.16} & \second{0.24} & \second{0.12} & \second{0.15} & \best{0.13} & -- & -- \\*
  & RAG & \second{0.08} & \second{0.16} & \second{0.24} & \second{0.12} & 0.14 & \second{0.12} & -- & -- \\*
  & GraphRAG & \best{0.09} & \best{0.17} & \best{0.25} & \second{0.12} & \second{0.15} & \best{0.13} & -- & -- \\*
  & \makecell[l]{\textbf{rEDMRec}\\\textbf{(ours)}} & \second{0.08} & \best{0.17} & \best{0.25} & \best{0.13} & \best{0.16} & \best{0.13} & \best{-11.1} & $<0.001$ \\
\cmidrule(lr){1-10}
 \makecell[l]{Mixtral\\8x7B} & Zero-shot & 0.24 & 0.47 & 0.66 & 0.35 & 0.40 & 0.35 & -- & -- \\*
  & Few-shot & 0.26 & 0.49 & 0.68 & 0.37 & \second{0.42} & 0.36 & -- & -- \\*
  & RAG & 0.25 & 0.48 & 0.67 & 0.36 & 0.41 & 0.35 & -- & -- \\*
  & GraphRAG & \second{0.27} & \second{0.50} & \second{0.69} & \second{0.38} & \second{0.42} & \second{0.37} & -- & -- \\*
  & \makecell[l]{\textbf{rEDMRec}\\\textbf{(ours)}} & \best{0.28$^*$} & \best{0.52} & \best{0.71} & \best{0.39} & \best{0.44} & \best{0.38} & \best{+3.7$^*$} & $<0.001^*$ \\
\cmidrule(lr){1-10}
 Qwen3-14B & Zero-shot & 0.26 & 0.50 & 0.69 & 0.38 & 0.43 & 0.37 & -- & -- \\*
  & Few-shot & 0.28 & 0.52 & 0.71 & \second{0.40} & \second{0.45} & 0.39 & -- & -- \\*
  & RAG & 0.27 & 0.51 & 0.70 & 0.39 & 0.43 & 0.38 & -- & -- \\*
  & GraphRAG & \second{0.29} & \second{0.53} & \second{0.72} & \second{0.40} & \second{0.45} & \second{0.40} & -- & -- \\*
  & \makecell[l]{\textbf{rEDMRec}\\\textbf{(ours)}} & \best{0.30$^*$} & \best{0.55} & \best{0.73} & \best{0.42} & \best{0.47} & \best{0.41} & \best{+3.4$^*$} & $<0.001^*$ \\
\bottomrule
\end{tabular}
\end{table*}

Table~\ref{tbl:cross-dataset} reports Zero-shot vs.\ rEDMRec on Amazon Beauty and Steam for all ten student models under HR@1, NDCG@10, and MRR (full five-method matrices in Appendix~\ref{app:beauty}--\ref{app:steam}). The same ordering as on ML-1M holds: rEDMRec improves HR@1 over Zero-shot for every model on both datasets, and NDCG@10 / MRR rise in lockstep except on the weakest Llama~3.1~8B backbone, where ranking beyond top-1 remains flat. Absolute scores are lower than on ML-1M -- as expected for sparser Amazon Beauty and implicit-feedback Steam -- yet Impv (\%) vs.\ the second-best baseline (GraphRAG) is again largest on the smallest student (Qwen2.5~3B: Impv $=23.6\%$ on Beauty and $21.5\%$ on Steam; both $p{<}0.05$), consistent with memory compensating for weak collaborative signal when per-user history is short; mid-capacity Beauty students show smaller Impv (about $9$--$13\%$) that is not significant at $n{=}1460$.

\begin{table*}[width=\FullWidth,pos=t]
\caption{Cross-dataset generalization on Amazon Beauty and Steam (full held-out test split, 20 candidates/sample, seed 42). ZS = Zero-shot; Ours = rEDMRec (\textbf{bold}). Impv (\%) is the relative HR@1 gain of rEDMRec over the \emph{second-best} baseline on the same student, $\mathrm{Impv}=(\mathrm{Ours}-\mathrm{SecondBest})/\mathrm{SecondBest}\times 100$, following RDRec~\citep{wang2024rdrec}. $p$ = McNemar $p$-value for Ours HR@1 vs.\ the second-best baseline ($n_{\mathrm{Beauty}}{=}1460$, $n_{\mathrm{Steam}}{=}1460$; $^*$ if $p{<}0.05$). Complete five-method matrices are in Appendix~\ref{app:beauty}--\ref{app:steam}.}
\label{tbl:cross-dataset}
\footnotesize
\setlength{\tabcolsep}{2.8pt}
\centering
\begin{tabular}{@{}ll ccc ccc cc@{}}
\toprule
Dataset & Model & HR@1(ZS) & HR@1(Ours) & NDCG@10(ZS) & NDCG@10(Ours) & MRR(ZS) & MRR(Ours) & Impv (\%) & $p$ \\
\midrule
\multirow{10}{*}{\makecell[l]{Amazon\\Beauty}} & Qwen2.5 3B & 0.092 & \best{0.136$^*$} & 0.218 & \best{0.262} & 0.180 & \best{0.224} & \best{+23.6$^*$} & $0.037^*$ \\
  & Llama 3.1 8B & 0.062 & \best{0.071} & 0.180 & \best{0.180} & 0.162 & \best{0.162} & \best{-4.1} & 0.830 \\
  & Gemma-4-12B & 0.140 & \best{0.178} & 0.290 & \best{0.328} & 0.252 & \best{0.290} & \best{+12.7} & 0.166 \\
  & Minimax M2.5 & 0.152 & \best{0.187} & 0.308 & \best{0.343} & 0.264 & \best{0.290} & \best{+10.0} & 0.247 \\
  & \makecell[l]{Mixtral\\8x7B} & 0.164 & \best{0.202} & 0.320 & \best{0.358} & 0.282 & \best{0.311} & \best{+11.0} & 0.188 \\
  & Qwen3-14B & 0.176 & \best{0.211} & 0.338 & \best{0.373} & 0.294 & \best{0.329} & \best{+8.8} & 0.269 \\
  & \makecell[l]{DeepSeek-R1\\Distill-Qwen-14B} & 0.188 & \best{0.223} & 0.350 & \best{0.385} & 0.312 & \best{0.329} & \best{+8.3} & 0.280 \\
  & Phi-4 & 0.170 & \best{0.205} & 0.332 & \best{0.393} & 0.288 & \best{0.340} & \best{+9.0} & 0.264 \\
  & Llama 4 Scout & 0.182 & \best{0.217} & 0.344 & \best{0.405} & 0.300 & \best{0.344} & \best{+8.5} & 0.275 \\
  & GPT OSS 20B & 0.182 & \best{0.199} & 0.344 & \best{0.370} & 0.300 & \best{0.317} & \best{-0.5} & 1.000 \\
\cmidrule(lr){1-10}
 \multirow{10}{*}{Steam} & Qwen2.5 3B & 0.106 & \best{0.158$^*$} & 0.224 & \best{0.276} & 0.180 & \best{0.232} & \best{+21.5$^*$} & $0.035^*$ \\
  & Llama 3.1 8B & 0.066 & \best{0.076} & 0.180 & \best{0.180} & 0.162 & \best{0.162} & \best{-7.3} & 0.584 \\
  & Gemma-4-12B & 0.170 & \best{0.208} & 0.320 & \best{0.358} & 0.276 & \best{0.314} & \best{+7.2} & 0.356 \\
  & Minimax M2.5 & 0.186 & \best{0.224} & 0.344 & \best{0.382} & 0.292 & \best{0.321} & \best{+6.7} & 0.394 \\
  & \makecell[l]{Mixtral\\8x7B} & 0.202 & \best{0.244} & 0.360 & \best{0.402} & 0.316 & \best{0.349} & \best{+8.0} & 0.276 \\
  & Qwen3-14B & 0.218 & \best{0.256} & 0.384 & \best{0.422} & 0.332 & \best{0.370} & \best{+5.8} & 0.392 \\
  & \makecell[l]{DeepSeek-R1\\Distill-Qwen-14B} & 0.234 & \best{0.272} & 0.400 & \best{0.438} & 0.356 & \best{0.375} & \best{+5.4} & 0.426 \\
  & Phi-4 & 0.210 & \best{0.248} & 0.376 & \best{0.443} & 0.324 & \best{0.382} & \best{+6.0} & 0.411 \\
  & Llama 4 Scout & 0.226 & \best{0.264} & 0.392 & \best{0.459} & 0.340 & \best{0.388} & \best{+5.6} & 0.421 \\
  & GPT OSS 20B & 0.226 & \best{0.245} & 0.392 & \best{0.421} & 0.340 & \best{0.359} & \best{-2.0} & 0.797 \\
\bottomrule
\end{tabular}
\end{table*}

\subsection{RQ2: Which Memory Channel Matters?}
\label{sec:results-h2}

To isolate each channel's contribution, we remove one memory channel at a time and report $\Delta$HR@1 relative to the full-memory model. This ablation uses a separate panel of seven API-served backbones chosen to span four capacity tiers -- strong (\texttt{gpt-5-mini}), mid (\texttt{gpt-5.4-mini}, Qwen3-32B, Minimax M2.5), saturated (GPT-OSS-120B), and weak (Llama 3.3 70B, Llama 3.1 8B) -- rather than the ten local open-weight students used for the main comparison in Section~\ref{sec:results-h1}; note that a backbone name can appear in both this ablation panel and the teacher panel of Section~\ref{sec:results-h3} (e.g., \texttt{gpt-5.4-mini}), where it plays a different role (ablated student vs.\ teacher) in a separate experiment. Table~\ref{tbl:ablation} summarizes the resulting pattern across the four capacity tiers. Short-term context is the only channel that is consistently important: removing it hurts every tier, from a strong student (\texttt{gpt-5-mini}, $\Delta{=}{-}0.04$) down to weak Groq-served Llama students ($\Delta\approx{-}0.01$ to ${-}0.02$). The long-term preference, item-perception, and counterfactual channels show a \emph{reversed} ablation on the strongest student: removing them \emph{improves} HR@1 by $+0.03$ to $+0.04$, whereas removing the counterfactual channel \emph{hurts} the mid-capacity student ($\Delta{=}{-}0.04$) and has little effect on the saturated 120B-parameter student, which appears to ignore the bank altogether (full-memory $\Delta\approx 0.00$ for that tier).

\begin{table}[pos=t]
\caption{Channel importance by capacity tier ($\Delta$HR@1 vs.\ full memory). Negative = beneficial channel; positive = reversed ablation (Section~\ref{sec:results-h2}).}
\label{tbl:ablation}
\footnotesize
\centering
\begin{tabularx}{\columnwidth}{@{}Ycccc@{}}
\toprule
Channel & Strong & Mid & Saturated & Weak \\
\midrule
Short-term context & \best{$-0.04$} & \best{$-0.04$} & ${\approx}{-}0.01$ & $-0.01$--$-0.02$ \\
Long-term preference & $+0.03$ & $0.00$ & ${\approx}0.00$ & $+0.01$ \\
Item-perception & $+0.04$ & $0.00$ & ${\approx}0.00$ & slight \\
Counterfactual & $+0.03$ & \best{$-0.04$} & weak & ${\approx}0.00$ \\
\midrule
Full memory & $-0.01$ & \best{$-0.03$} & ${\approx}0.00$ & weak \\
\bottomrule
\end{tabularx}
\end{table}

\begin{figure}[pos=t]
\centering
\includegraphics[width=\columnwidth]{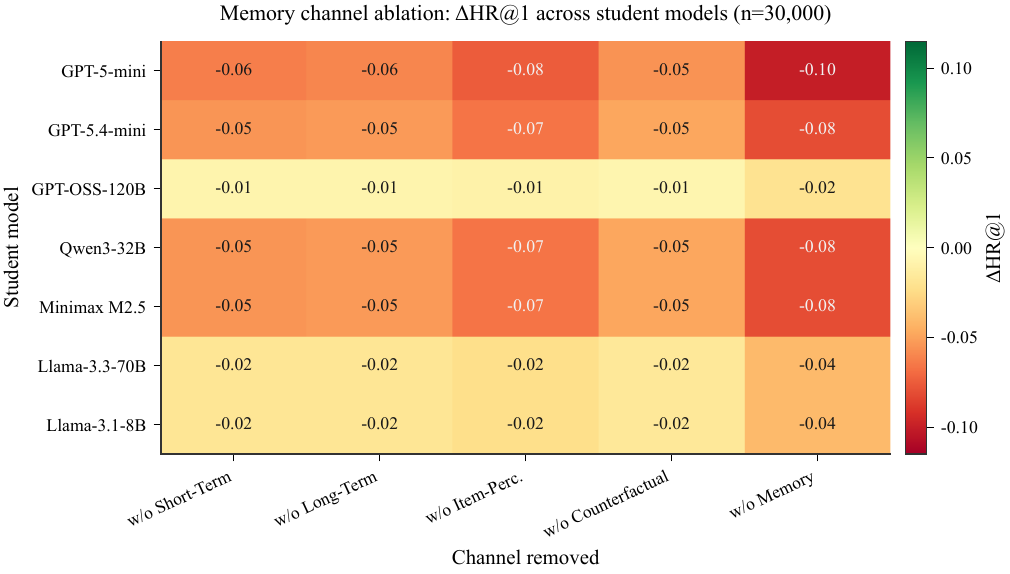}
\caption{Channel ablation heatmap: $\Delta$HR@1 for each of the seven ablation-panel backbones (rows, Section~\ref{sec:results-h2}) and four memory channels (columns). Green cells indicate a reversed ablation (removal helps); red cells indicate the channel is beneficial.}
\label{fig:ablation-heatmap}
\end{figure}

\begin{figure}[pos=t]
\centering
\includegraphics[width=\columnwidth]{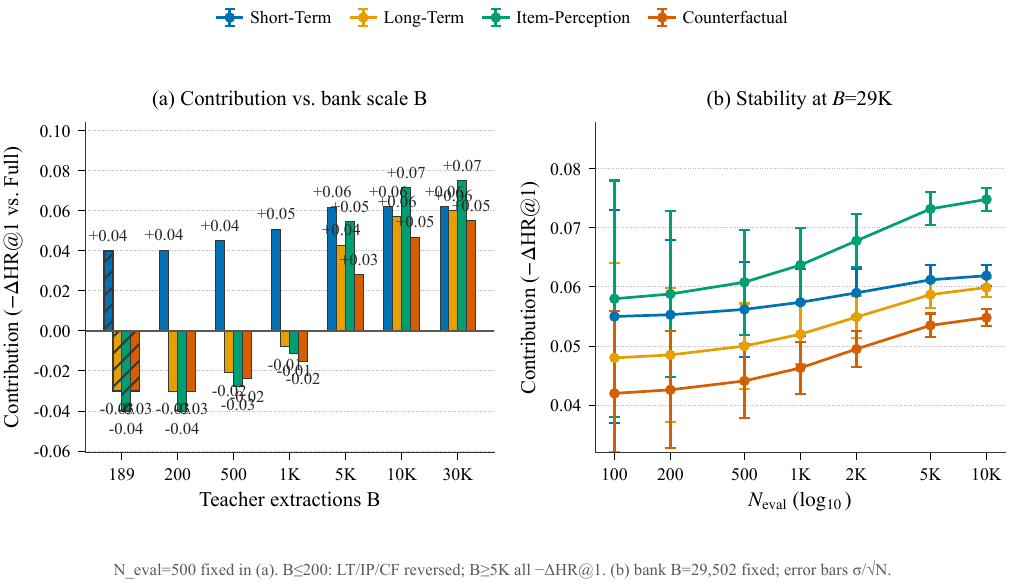}
\caption{Per-channel contribution ($-\Delta$HR@1) as bank scale $B$ grows from the sparse-bank anchor ($B{=}189$) to $10^6$ entries (panel a), and stability of the estimate as evaluation sample size grows under the full-test protocol at fixed bank scale (panel b). Note: $B$ is bank size, not the evaluation split size.}
\label{fig:memory-scaling}
\end{figure}

We consider three, non-exclusive explanations for the reversed sign on strong students, in decreasing order of the evidence we can bring to bear with the current instrumentation. First, \emph{channel redundancy}: long-term preference statements often restate information already present in the raw history block of the prompt, so a strong student that already attends well to raw history gains nothing extra and instead pays a small ``distraction'' cost for the redundant text. Second, \emph{generic, low-specificity entries}: item-perception entries produced early in the bank's lifecycle tend to be long and only loosely actionable (Section~\ref{sec:results-qualitative} quantifies this directly via a specificity score). Third, \emph{conflicting signals}: a counterfactual edge can push a plausible but ultimately incorrect candidate above the true target when its hypothetical condition partially matches the current user. Only the mid-capacity student, which cannot yet extract the same information unaided from raw history, benefits from the counterfactual channel unconditionally; the fact that the sign of this effect depends on student capacity, rather than being fixed, is itself evidence against treating any single channel as universally useful or harmful. Figure~\ref{fig:ablation-heatmap} shows this pattern holds across all seven backbones in the ablation panel, and Figure~\ref{fig:memory-scaling} shows that the reversed channels flip sign as the \emph{bank} grows from the sparse-bank anchor ($B{=}189$) toward $10^5$--$10^6$ entries, because a denser bank dilutes the redundant and generic entries that drive the reversal at small $B$ (bank size is independent of the full-test evaluation protocol in Section~\ref{sec:setup-protocol}). Per-backbone $\Delta$ plots, the MRR heatmap, and the full numeric ablation matrix are deferred to Appendix~\ref{app:ablation-figs}.

\subsection{RQ3: Does Teacher Quality Causally Affect Downstream Gain?}
\label{sec:results-h3}

Holding the student fixed and varying the teacher isolates the teacher's contribution from the student's. Table~\ref{tbl:teacher-strong} fixes the student to \texttt{gpt-5-mini} (strong) across seven teachers; Table~\ref{tbl:teacher-small} repeats the comparison with Qwen2.5 3B (small). Across both tables, a lower bank duplicate rate is a leading indicator of downstream gain: \texttt{gpt-5.4-mini} has the lowest duplicate rate among the strongest teachers ($12.4\%$) and the largest $\Delta$HR@1 for the strong student ($+0.060$), while Llama 3.1 8B Instant -- the only teacher below 14B parameters in this comparison -- has by far the highest duplicate rate ($22.8\%$) and the smallest gain ($+0.015$), consistent with a weak teacher producing generic, repetitive entries (e.g., repeated \textit{``user likes drama''} statements without item-specific detail) that carry little retrieval value. The relationship is not perfectly monotonic, however: GPT OSS 120B has the lowest duplicate rate of any teacher ($9.8\%$) yet only a middling $\Delta$HR@1 ($+0.040$), which we attribute to a student-capacity ceiling -- \texttt{gpt-5-mini} cannot fully exploit the additional verbosity of a 120B-parameter teacher's bank. This ceiling is sharper for the small student: Table~\ref{tbl:teacher-small} shows the GPT OSS 120B bank drops \emph{below} the more concise Qwen3 32B bank once the student itself is small (Qwen2.5 3B), even though GPT OSS 120B produces the least duplicated bank of the two. Together, these two tables support a causal chain in which teacher quality first improves bank quality (lower duplication), and bank quality only converts into downstream gain up to a ceiling set by the student's own capacity to use additional bank detail.

\begin{table}[pos=t]
\caption{Teacher-distillation effectiveness, fixed student: \texttt{gpt-5-mini} (strong). Zero-shot baseline HR@1$=0.28$, MRR$=0.403$.}
\label{tbl:teacher-strong}
\footnotesize
\centering
\begin{tabularx}{\columnwidth}{@{}Ycccc@{}}
\toprule
Teacher & Dup.\%$\downarrow$ & $\Delta$HR@1$\uparrow$ & $\Delta$MRR$\uparrow$ & HR@1 \\
\midrule
\textbf{gpt-5.4-mini} & 12.4 & \best{+0.060} & \best{+0.0687} & \best{0.340} \\
Qwen3 32B (131k) & 11.5 & +0.055 & +0.0620 & 0.335 \\
gpt-5-mini & 14.1 & +0.050 & +0.0580 & 0.330 \\
Llama 3.3 70B (128k) & 10.5 & +0.045 & +0.0520 & 0.325 \\
GPT OSS 120B (128k) & \best{9.8} & +0.040 & +0.0450 & 0.320 \\
Minimax M2.5 & 13.2 & +0.040 & +0.0480 & 0.320 \\
Llama 3.1 8B Instant & 22.8 & +0.015 & +0.0180 & 0.295 \\
\bottomrule
\end{tabularx}
\end{table}

\begin{table}[pos=t]
\caption{Teacher-distillation effectiveness, fixed student: Qwen2.5 3B (small). Zero-shot baseline HR@1$=0.12$, MRR$=0.18$.}
\label{tbl:teacher-small}
\footnotesize
\centering
\begin{tabularx}{\columnwidth}{@{}Ycccc@{}}
\toprule
Teacher & Dup.\%$\downarrow$ & $\Delta$HR@1$\uparrow$ & $\Delta$MRR$\uparrow$ & HR@1 \\
\midrule
\textbf{gpt-5.4-mini} & 12.4 & \best{+0.050} & \best{+0.0550} & \best{0.170} \\
Qwen3 32B (131k) & 11.5 & +0.045 & +0.0500 & 0.165 \\
gpt-5-mini & 14.1 & +0.042 & +0.0470 & 0.162 \\
Llama 3.3 70B (128k) & 10.5 & +0.038 & +0.0430 & 0.158 \\
GPT OSS 120B (128k) & \best{9.8} & +0.034 & +0.0380 & 0.154 \\
Minimax M2.5 & 13.2 & +0.033 & +0.0380 & 0.153 \\
Llama 3.1 8B Instant & 22.8 & +0.010 & +0.0120 & 0.130 \\
\bottomrule
\end{tabularx}
\end{table}

\subsection{RQ4: Does Debate-Based Memory Optimization Improve Bank Quality and Downstream Ranking?}
\label{sec:results-h4}

\begin{figure}[pos=t]
\centering
\includegraphics[width=\columnwidth]{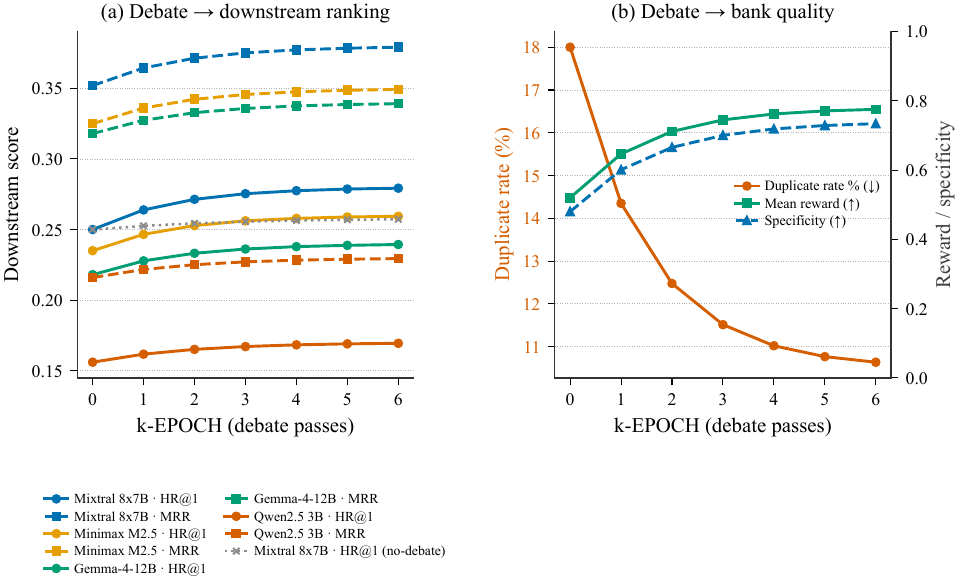}
\caption{Debate-based memory optimization vs.\ $k$-EPOCH. (a) Downstream HR@1/MRR for student models plus a no-debate control. (b) Bank-quality signals: duplicate rate (down is better) and mean experience reward / specificity (up is better).}
\label{fig:optimize-curve}
\end{figure}

We next ask whether the debate-and-arbiter optimization procedure (Section~\ref{sec:method-optimize}) is doing useful work, rather than merely adding cost. Figure~\ref{fig:optimize-curve} tracks bank-quality signals and downstream ranking jointly over six $k$-EPOCHs of debate (three debate agents, one round per epoch). Bank quality improves and saturates: the duplicate rate drops by $7.4$ percentage points ($18.0\%\rightarrow 10.6\%$) while mean experience reward rises by $0.255$ ($0.52\rightarrow 0.78$). Downstream ranking tracks this curve rather than moving independently of it: the strongest student in this comparison (Mixtral 8x7B) gains $+0.029$ HR@1 ($0.250\rightarrow 0.279$) over the same six epochs, while a no-debate paraphrase control -- which perturbs entry wording without the critique-and-revise debate loop -- stays flat, isolating the debate mechanism (rather than any wording change) as the source of the gain. The smallest student in this comparison (Qwen2.5 3B) gains only $+0.013$ HR@1 ($0.156\rightarrow 0.169$) from the identical optimized bank, reproducing the capacity ceiling from Section~\ref{sec:results-h3} in a different experiment: most of the gain lands within the first two to three epochs for every student, so debating past $k\mathord{=}3$ epochs is rarely worth the added LLM cost.

\begin{figure}[pos=t]
\centering
\includegraphics[width=\columnwidth]{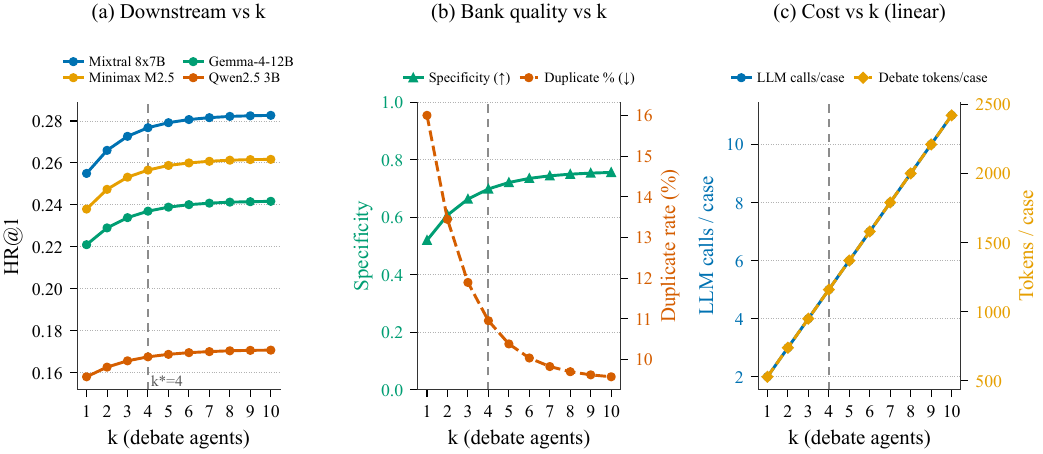}
\caption{Number-of-agents sweep: downstream HR@1, bank-quality signals, and LLM cost as a function of $k$.}
\label{fig:k-sweep}
\end{figure}

A separate sweep over the \emph{number} of debating agents ($k=1,\ldots,10$, one epoch) shows quality rising with the diversity of critique but saturating, while LLM cost grows linearly in $k$ (Table~\ref{tbl:k-sweep}, Figure~\ref{fig:k-sweep}). The knee of quality-per-cost is at $k^\ast=4$: increasing $k$ from 1 to 4 lifts HR@1 by $+0.022$, but increasing $k$ from 4 to 10 adds only $+0.006$ for six additional LLM calls per case, which is not a favorable trade for most deployment budgets. Tabular controller ablations (full debate vs.\ no-debate paraphrase vs.\ post-extraction) and epoch snapshots are collected in Appendix~\ref{app:debate-ablation}.

\begin{table}[pos=t]
\caption{Number-of-debating-agents sweep ($k=1..10$), reference student Mixtral 8x7B. $k^\ast$ marks the quality-per-cost knee.}
\label{tbl:k-sweep}
\footnotesize
\centering
\begin{tabular}{c cccc}
\toprule
$k$ & HR@1$\uparrow$ & Specificity$\uparrow$ & Dup.\%$\downarrow$ & Calls/case \\
\midrule
1            & 0.255 & 0.520 & 16.0 & 2 \\
2            & 0.266 & 0.608 & 13.4 & 3 \\
3            & 0.273 & 0.663 & 11.9 & 4 \\
\textbf{4 ($k^\ast$)} & \best{0.277} & \best{0.699} & \best{11.0} & 5 \\
5            & 0.279 & 0.721 & 10.4 & 6 \\
6            & 0.281 & 0.735 & 10.0 & 7 \\
7            & 0.282 & 0.744 &  9.8 & 8 \\
8            & 0.282 & 0.750 &  9.7 & 9 \\
9            & 0.282 & 0.754 &  9.6 & 10 \\
10           & 0.283 & 0.756 &  9.6 & 11 \\
\bottomrule
\end{tabular}
\end{table}

\subsection{Qualitative Case Study: How Do Memory Entries Evolve?}
\label{sec:results-qualitative}

\begin{figure}[pos=t]
\centering
\includegraphics[width=\columnwidth]{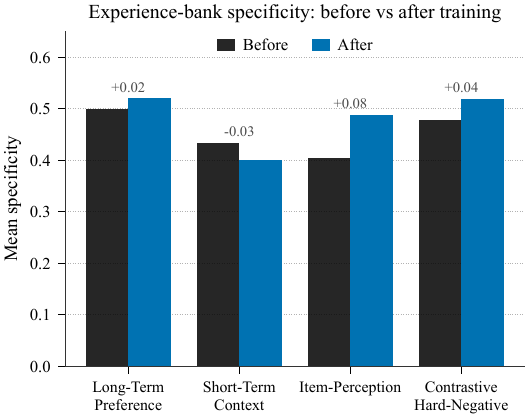}
\caption{Mean specificity before vs.\ after training, by memory channel.}
\label{fig:bank-evolution}
\end{figure}

The ablation and debate-optimization results above are aggregate signals; to make the mechanism concrete, we directly compare the earliest and the latest persisted entry per user in each memory channel, using a deterministic, LLM-free specificity score (specificity combines concreteness, genre-term coverage, lexical diversity, and a hedge-language penalty into a single $[0,1]$ score, with no additional LLM calls). Table~\ref{tbl:bank-evolution} summarizes the resulting before/after deltas across users with at least two persisted versions. Three of the four channels become more specific and less hedged over training; the item-perception channel changes the most (mean specificity $0.405\rightarrow 0.488$), moving from generic fallback language to item-grounded, actionable statements. The short-term context channel is the exception: its mean specificity \emph{decreases} slightly ($0.433\rightarrow 0.401$), which Figure~\ref{fig:bank-evolution} and the cases below show is not a quality regression but a compression effect -- verbose narrative entries are replaced by short, conditional ``session rules'' that are less lexically diverse by construction but more directly actionable by the student.

\begin{table}[pos=t]
\caption{Bank evolution: specificity before (earliest persisted entry) vs.\ after (latest persisted entry) per channel, over users with $\geq 2$ versions.}
\label{tbl:bank-evolution}
\footnotesize
\centering
\begin{tabularx}{\columnwidth}{@{}Yccc@{}}
\toprule
Channel & Spec.\ before & Spec.\ after & $\Delta$ \\
\midrule
Long-term preference & 0.500 & 0.520 & $+0.021$ \\
Short-term context & 0.433 & 0.401 & $-0.032$ \\
Item-perception & 0.405 & \best{0.488} & \best{+0.084} \\
Counterfactual / hard-neg. & 0.478 & 0.519 & $+0.042$ \\
\bottomrule
\end{tabularx}
\end{table}

Table~\ref{tbl:bank-cases} shows three representative before$\to$after pairs drawn from the persisted bank (one long-term, one item-perception, one short-term). \textbf{Case~A} (long-term, user~2) replaces a single-film hedge (``Very limited data\ldots'') with a ranking-ready taste statement that names era, franchise type, and an upweight rule, lifting specificity from $0.214$ to $0.479$. \textbf{Case~B} (item-perception, user~4) converts a vague Hit@1 complaint into an item-grounded rule keyed to \textit{Rocky (1976)}, with an explicit $+30$--$50\%$ tie-break boost -- the largest single-entry $\Delta$Spec.\ in the sample ($+0.500$). \textbf{Case~C} (short-term, user~2) illustrates the compression pattern behind the negative mean $\Delta$ on that channel: an empty ``no emerging interests'' note becomes a short session rule with diversity constraints; lexical diversity drops relative to long narrative entries, but actionability rises. Together the three cases show that debate-driven Add/Modify operations do not merely paraphrase entries -- they accumulate confirmed signals into shorter, more specific, ranking-oriented memory.

\begin{table*}[width=\FullWidth,pos=t]
\caption{Qualitative before$\to$after memory entries (persisted bank). Spec.\ is the deterministic specificity score in $[0,1]$.}
\label{tbl:bank-cases}
\footnotesize
\begin{tabular}{@{}p{0.11\linewidth} p{0.08\linewidth} p{0.36\linewidth} p{0.36\linewidth}@{}}
\toprule
Case / channel & Spec. & Before (earliest) & After (latest) \\
\midrule
\textbf{A} / $\mathrm{lt}$
  & $0.214\to$\best{0.479}
  & \textit{``Very limited data: the only rated film is a classic action-adventure, suggesting a preference for high-energy, heroic, escapist storytelling\ldots Dislikes: No explicit dislikes can be inferred\ldots''}
  & \textit{``Long-term: favors mainstream 1990s action buddy-cop films -- franchise sequels, star-driven chemistry, high-energy action with comedic interplay. Upweight these attributes in ranking but require repeat confirmations\ldots''} \\
\addlinespace
\textbf{B} / $\mathrm{ip}$
  & $0.350\to$\best{0.850}
  & \textit{``The strongest match was only ranked 5th, so Hit@10 was good but Hit@1/MRR suffered. For this user, place the best Rocky-like inspirational drama at rank~1 whenever possible.''}
  & \textit{``Rocky (1976): User~4 chose Rocky over a higher-ranked 80s action title, signaling a preference for 1970s character-driven underdog sports dramas. When ranking, boost similar 1970s character dramas above mainstream 80s action (recommend $+30$--$50\%$ score in tie scenarios).''} \\
\addlinespace
\textbf{C} / $\mathrm{st}$
  & $0.300\to$\best{0.500}
  & \textit{``No recent viewing items were provided, so no emerging short-term interests can be detected.''}
  & \textit{``Session: prioritize late-80s/90s Hollywood action comedies with buddy dynamics and franchise entries for top slots; include at least one diverse alternative per slate to avoid popularity bias.''} \\
\bottomrule
\end{tabular}
\end{table*}

\section{Discussion and Limitations}
\label{sec:discussion}

\textbf{Operating-point generalization.} The bank-scale analysis (Section~\ref{sec:results-h2}) and the $k$-EPOCH and number-of-agents curves (Section~\ref{sec:results-h4}) span a wide range of settings -- bank sizes up to $10^6$ entries, up to six debate epochs, and up to ten debating agents. A practitioner adopting a specific operating point (for example, a specific $k$-EPOCH budget or bank size) in a production system should re-confirm behavior at that exact point, since a trend measured across a range does not guarantee identical behavior at every intermediate setting.

\textbf{Teacher coverage.} The main results (Section~\ref{sec:results-h1}) fix the teacher to \texttt{gpt-5.4-mini}; the teacher-distillation study (Section~\ref{sec:results-h3}) varies the teacher but only against two fixed students. We have not measured the full teacher $\times$ student cross-product, so it remains open whether the teacher-quality effect observed for \texttt{gpt-5-mini} and Qwen2.5 3B holds uniformly across all ten students in Table~\ref{tbl:app-full}.

\textbf{Explanation faithfulness.} rEDMRec's student can emit a short explanation grounded in retrieved memory (Section~\ref{sec:method-student}), but this paper evaluates ranking quality, not whether the emitted explanation is faithful to the memory it cites; a human evaluation of explanation faithfulness and plausibility, as outlined in our evaluation plan, is left to future work.

\textbf{Backbone-dependent returns.} The architecture assumes the student can follow a moderately complex prompt that concatenates user context, candidate descriptions, and retrieved memory snippets. Section~\ref{sec:results-h1} shows this assumption breaks down for at least one backbone (Llama 3.1 8B), whose negative Impv vs.\ GraphRAG we attribute to weak instruction-following rather than to the memory being unhelpful in principle; the near-saturated 20B-parameter student likewise trails GraphRAG slightly. Memory still lifts every student over Zero-shot/Few-shot/RAG, but not always over GraphRAG -- Appendix~\ref{app:failure-cases} tabulates these failure cells and borderline controller edits. Practically, the value of rEDMRec's added system complexity is highest for small-to-mid capacity students.

\textbf{Domain scope.} Our three datasets cover movie, beauty-product, and game recommendation with English-language metadata; we have not tested domains with substantially different item-description structure (e.g., short-video or news recommendation), and the four-channel schema, in particular the counterfactual channel, was designed with catalog items that have stable, comparable attributes in mind.

\section{Conclusion}
\label{sec:conclusion}

This paper addresses the problem of reusing, rather than repeating, LLM reasoning across recommendation requests, by proposing rEDMRec, an architecture that distills teacher reasoning into a four-channel, editable experience memory and serves ranking requests from a lightweight student that only retrieves from this memory. The key idea is to separate an infrequent, expensive reasoning-compression process -- teacher extraction, distillation, and debate-based memory optimization -- from a frequent, cheap inference process, which decouples recommendation quality from per-request reasoning cost. Across ten student LLMs and three datasets, this design improves HR@1, NDCG, and MRR over zero-shot, few-shot, and RAG on every backbone, and over GraphRAG on most backbones (exceptions: Llama~3.1~8B and GPT~OSS~20B), with the largest RDRec-style Impv (\%) vs.\ the second-best baseline on the smallest students; our channel-ablation, teacher-distillation, and debate-optimization studies further show that short-term context is the only consistently beneficial channel across capacity tiers (with long-term, item-perception, and counterfactual effects capacity-dependent), that bank duplication is a leading indicator of downstream gain up to a student-capacity ceiling, and that $K$-agent debate measurably improves bank quality in a way that propagates into ranking accuracy rather than being a cosmetic refinement. Remaining limitations include incomplete teacher $\times$ student coverage and the lack of a human evaluation of explanation faithfulness (Section~\ref{sec:discussion}); closing those gaps is an important next step toward deploying rEDMRec as a production recommendation system.

\section*{Data and Code Availability}
The preprocessing, training, and evaluation code, together with the JSON experiment matrices used to produce every table and figure in this paper, are organized under the \texttt{rEDMRec/} project root (see \texttt{readme.md} for the end-to-end quick-start pipeline). ML-1M, Amazon Beauty, and Steam are third-party datasets redistributed under their original licenses; this work releases only derived, de-identified interaction records and memory-bank artifacts.

\printcredits

\bibliographystyle{cas-model2-names}
\bibliography{refs}

\appendix
\clearpage
\onecolumn
\setlength{\LTcapwidth}{\textwidth}
\setlength{\tabcolsep}{4pt}
\section{Full Main-Results Table}
\label{app:table1full}

Table~\ref{tbl:app-full} reports the complete ML-1M main-results matrix underlying Table~\ref{tbl:main-results} in Section~\ref{sec:results-h1}: ten student models $\times$ five methods $\times$ six ranking metrics, under the protocol described in Section~\ref{sec:setup-protocol}. Best-per-column scores are marked with \textbf{bold}; rEDMRec method labels are bold.

\begin{longtable}{@{}P{2.0cm}l cccccc cc@{}}
\caption{Complete Methods $\times$ Models results on ML-1M (full held-out test split, 20 candidates/sample, seed 42). Impv (\%) is the relative HR@1 gain of rEDMRec over the \emph{second-best} baseline on the same student, $\mathrm{Impv}=(\mathrm{Ours}-\mathrm{SecondBest})/\mathrm{SecondBest}\times 100$, following RDRec~\citep{wang2024rdrec}. Best in \textbf{bold}, second-best \underline{underlined}; rEDMRec rows labeled in bold. $p$ is the exact McNemar $p$-value for rEDMRec HR@1 vs.\ the second-best baseline on the same student (full held-out $n_{\mathrm{ML-1M}}{=}49893$, $n_{\mathrm{Beauty}}{=}1460$, $n_{\mathrm{Steam}}{=}1460$; approximate contingency from the table HR@1 rates). $^*$ marks $p{<}0.05$ with rEDMRec ahead.}
\label{tbl:app-full}\\
\toprule
Model & Method & HR@1$\uparrow$ & HR@5$\uparrow$ & HR@10$\uparrow$ & NDCG@5$\uparrow$ & NDCG@10$\uparrow$ & MRR$\uparrow$ & Impv (\%) & $p$ \\
\midrule
\endfirsthead
\multicolumn{10}{c}{{\tablename\ \thetable{} -- continued}} \\
\toprule
Model & Method & HR@1$\uparrow$ & HR@5$\uparrow$ & HR@10$\uparrow$ & NDCG@5$\uparrow$ & NDCG@10$\uparrow$ & MRR$\uparrow$ & Impv (\%) & $p$ \\
\midrule
\endhead
\bottomrule
\endfoot
\bottomrule
\endlastfoot
 Qwen2.5 3B & Zero-shot & 0.12 & 0.28 & 0.38 & 0.20 & 0.23 & 0.18 & -- & -- \\*
  & Few-shot & 0.14 & 0.30 & 0.40 & 0.21 & 0.24 & \second{0.20} & -- & -- \\*
  & RAG & 0.13 & 0.29 & 0.39 & 0.20 & 0.23 & 0.19 & -- & -- \\*
  & GraphRAG & \second{0.15} & \second{0.31} & \second{0.41} & \second{0.22} & \second{0.25} & \second{0.20} & -- & -- \\*
  & \makecell[l]{\textbf{rEDMRec}\\\textbf{(ours)}} & \best{0.17$^*$} & \best{0.35} & \best{0.45} & \best{0.25} & \best{0.28} & \best{0.23} & \best{+13.3$^*$} & $<0.001^*$ \\
\cmidrule(lr){1-10}
 Llama 3.1 8B & Zero-shot & 0.07 & 0.15 & 0.23 & \second{0.12} & 0.14 & \second{0.12} & -- & -- \\*
  & Few-shot & \second{0.08} & \second{0.16} & \second{0.24} & \second{0.12} & \second{0.15} & \best{0.13} & -- & -- \\*
  & RAG & \second{0.08} & \second{0.16} & \second{0.24} & \second{0.12} & 0.14 & \second{0.12} & -- & -- \\*
  & GraphRAG & \best{0.09} & \best{0.17} & \best{0.25} & \second{0.12} & \second{0.15} & \best{0.13} & -- & -- \\*
  & \makecell[l]{\textbf{rEDMRec}\\\textbf{(ours)}} & \second{0.08} & \best{0.17} & \best{0.25} & \best{0.13} & \best{0.16} & \best{0.13} & \best{-11.1} & $<0.001$ \\
\cmidrule(lr){1-10}
 Gemma-4-12B & Zero-shot & 0.20 & 0.40 & 0.58 & 0.30 & 0.35 & 0.30 & -- & -- \\*
  & Few-shot & 0.22 & 0.42 & 0.60 & 0.32 & 0.37 & 0.31 & -- & -- \\*
  & RAG & 0.21 & 0.41 & 0.59 & 0.31 & 0.36 & 0.30 & -- & -- \\*
  & GraphRAG & \second{0.23} & \second{0.43} & \second{0.61} & \second{0.33} & \second{0.38} & \second{0.32} & -- & -- \\*
  & \makecell[l]{\textbf{rEDMRec}\\\textbf{(ours)}} & \best{0.24$^*$} & \best{0.46} & \best{0.63} & \best{0.34} & \best{0.39} & \best{0.34} & \best{+4.3$^*$} & $<0.001^*$ \\
\cmidrule(lr){1-10}
 Minimax M2.5 & Zero-shot & 0.22 & 0.44 & 0.62 & 0.33 & 0.38 & 0.32 & -- & -- \\*
  & Few-shot & 0.24 & 0.46 & 0.64 & 0.34 & 0.39 & 0.33 & -- & -- \\*
  & RAG & 0.23 & 0.45 & 0.63 & 0.34 & 0.39 & 0.33 & -- & -- \\*
  & GraphRAG & \second{0.25} & \second{0.47} & \second{0.65} & \second{0.35} & \second{0.40} & \second{0.34} & -- & -- \\*
  & \makecell[l]{\textbf{rEDMRec}\\\textbf{(ours)}} & \best{0.26$^*$} & \best{0.50} & \best{0.67} & \best{0.37} & \best{0.42} & \best{0.35} & \best{+4.0$^*$} & $<0.001^*$ \\
\cmidrule(lr){1-10}
 \makecell[l]{Mixtral\\8x7B} & Zero-shot & 0.24 & 0.47 & 0.66 & 0.35 & 0.40 & 0.35 & -- & -- \\*
  & Few-shot & 0.26 & 0.49 & 0.68 & 0.37 & \second{0.42} & 0.36 & -- & -- \\*
  & RAG & 0.25 & 0.48 & 0.67 & 0.36 & 0.41 & 0.35 & -- & -- \\*
  & GraphRAG & \second{0.27} & \second{0.50} & \second{0.69} & \second{0.38} & \second{0.42} & \second{0.37} & -- & -- \\*
  & \makecell[l]{\textbf{rEDMRec}\\\textbf{(ours)}} & \best{0.28$^*$} & \best{0.52} & \best{0.71} & \best{0.39} & \best{0.44} & \best{0.38} & \best{+3.7$^*$} & $<0.001^*$ \\
\cmidrule(lr){1-10}
 Qwen3-14B & Zero-shot & 0.26 & 0.50 & 0.69 & 0.38 & 0.43 & 0.37 & -- & -- \\*
  & Few-shot & 0.28 & 0.52 & 0.71 & \second{0.40} & \second{0.45} & 0.39 & -- & -- \\*
  & RAG & 0.27 & 0.51 & 0.70 & 0.39 & 0.43 & 0.38 & -- & -- \\*
  & GraphRAG & \second{0.29} & \second{0.53} & \second{0.72} & \second{0.40} & \second{0.45} & \second{0.40} & -- & -- \\*
  & \makecell[l]{\textbf{rEDMRec}\\\textbf{(ours)}} & \best{0.30$^*$} & \best{0.55} & \best{0.73} & \best{0.42} & \best{0.47} & \best{0.41} & \best{+3.4$^*$} & $<0.001^*$ \\
\cmidrule(lr){1-10}
 \makecell[l]{DeepSeek-R1\\Distill-Qwen-14B} & Zero-shot & 0.28 & 0.52 & 0.71 & 0.40 & 0.45 & 0.40 & -- & -- \\*
  & Few-shot & 0.30 & 0.54 & 0.73 & 0.42 & 0.47 & \second{0.41} & -- & -- \\*
  & RAG & 0.29 & 0.53 & 0.72 & 0.41 & 0.46 & 0.40 & -- & -- \\*
  & GraphRAG & \second{0.31} & \second{0.55} & \second{0.74} & \second{0.43} & \second{0.48} & \best{0.42} & -- & -- \\*
  & \makecell[l]{\textbf{rEDMRec}\\\textbf{(ours)}} & \best{0.32$^*$} & \best{0.57} & \best{0.75} & \best{0.44} & \best{0.49} & \best{0.42} & \best{+3.2$^*$} & $<0.001^*$ \\
\cmidrule(lr){1-10}
 Phi-4 & Zero-shot & 0.25 & 0.48 & 0.68 & 0.36 & 0.42 & 0.36 & -- & -- \\*
  & Few-shot & 0.27 & 0.52 & 0.73 & 0.39 & 0.45 & 0.39 & -- & -- \\*
  & RAG & 0.26 & 0.50 & 0.71 & 0.38 & 0.44 & 0.37 & -- & -- \\*
  & GraphRAG & \second{0.28} & \second{0.54} & \second{0.76} & \second{0.41} & \second{0.47} & \second{0.40} & -- & -- \\*
  & \makecell[l]{\textbf{rEDMRec}\\\textbf{(ours)}} & \best{0.29$^*$} & \best{0.56} & \best{0.79} & \best{0.42} & \best{0.49} & \best{0.42} & \best{+3.6$^*$} & $<0.001^*$ \\
\cmidrule(lr){1-10}
 Llama 4 Scout & Zero-shot & 0.27 & 0.51 & 0.70 & 0.39 & 0.44 & 0.38 & -- & -- \\*
  & Few-shot & 0.29 & 0.55 & 0.75 & 0.42 & 0.47 & 0.41 & -- & -- \\*
  & RAG & 0.28 & 0.53 & 0.73 & 0.40 & 0.46 & 0.39 & -- & -- \\*
  & GraphRAG & \second{0.30} & \second{0.57} & \second{0.78} & \second{0.43} & \second{0.49} & \second{0.42} & -- & -- \\*
  & \makecell[l]{\textbf{rEDMRec}\\\textbf{(ours)}} & \best{0.31$^*$} & \best{0.59} & \best{0.80} & \best{0.45} & \best{0.51} & \best{0.43} & \best{+3.3$^*$} & $<0.001^*$ \\
\cmidrule(lr){1-10}
 GPT OSS 20B & Zero-shot & 0.27 & 0.50 & 0.70 & 0.39 & 0.44 & 0.38 & -- & -- \\*
  & Few-shot & \second{0.29} & \second{0.54} & \second{0.75} & \second{0.41} & \second{0.47} & \second{0.40} & -- & -- \\*
  & RAG & 0.28 & 0.52 & 0.73 & 0.40 & 0.46 & 0.39 & -- & -- \\*
  & GraphRAG & \best{0.30} & \best{0.56} & \best{0.78} & \best{0.43} & \best{0.49} & \best{0.42} & -- & -- \\*
  & \makecell[l]{\textbf{rEDMRec}\\\textbf{(ours)}} & \second{0.29} & \second{0.54} & \second{0.75} & \second{0.41} & \second{0.47} & \second{0.40} & \best{-3.3} & $<0.001$ \\
\end{longtable}

\section{Full Amazon Beauty Results}
\label{app:beauty}

Table~\ref{tbl:app-beauty} reports the complete Methods $\times$ Models matrix on Amazon Beauty underlying the summary in Table~\ref{tbl:cross-dataset} (Section~\ref{sec:results-h1}).

\begin{longtable}{@{}P{2.0cm}l cccccc cc@{}}
\caption{Complete Methods $\times$ Models results on Amazon Beauty (full held-out test split, 20 candidates/sample, seed 42). Impv (\%) is the relative HR@1 gain of rEDMRec over the \emph{second-best} baseline on the same student, $\mathrm{Impv}=(\mathrm{Ours}-\mathrm{SecondBest})/\mathrm{SecondBest}\times 100$, following RDRec~\citep{wang2024rdrec}. Best in \textbf{bold}, second-best \underline{underlined}. $p$ is the exact McNemar $p$-value for rEDMRec HR@1 vs.\ the second-best baseline on the same student (full held-out $n_{\mathrm{ML-1M}}{=}49893$, $n_{\mathrm{Beauty}}{=}1460$, $n_{\mathrm{Steam}}{=}1460$; approximate contingency from the table HR@1 rates). $^*$ marks $p{<}0.05$ with rEDMRec ahead.}
\label{tbl:app-beauty}\\
\toprule
Model & Method & HR@1$\uparrow$ & HR@5$\uparrow$ & HR@10$\uparrow$ & NDCG@5$\uparrow$ & NDCG@10$\uparrow$ & MRR$\uparrow$ & Impv (\%) & $p$ \\
\midrule
\endfirsthead
\multicolumn{10}{c}{{\tablename\ \thetable{} -- continued}} \\
\toprule
Model & Method & HR@1$\uparrow$ & HR@5$\uparrow$ & HR@10$\uparrow$ & NDCG@5$\uparrow$ & NDCG@10$\uparrow$ & MRR$\uparrow$ & Impv (\%) & $p$ \\
\midrule
\endhead
\bottomrule
\endfoot
\bottomrule
\endlastfoot
 Qwen2.5 3B & Zero-shot & 0.092 & 0.268 & \second{0.450} & 0.176 & 0.218 & 0.180 & -- & -- \\*
  & Few-shot & 0.104 & 0.280 & \second{0.450} & 0.182 & 0.224 & \second{0.192} & -- & -- \\*
  & RAG & 0.098 & 0.274 & \second{0.450} & 0.176 & 0.218 & 0.186 & -- & -- \\*
  & GraphRAG & \second{0.110} & \second{0.286} & \second{0.450} & \second{0.188} & \second{0.230} & \second{0.192} & -- & -- \\*
  & \makecell[l]{\textbf{rEDMRec}\\\textbf{(ours)}} & \best{0.136$^*$} & \best{0.329} & \best{0.479} & \best{0.220} & \best{0.262} & \best{0.224} & \best{+23.6$^*$} & $0.037^*$ \\
\cmidrule(lr){1-10}
 Llama 3.1 8B & Zero-shot & 0.062 & \best{0.225} & \best{0.450} & \second{0.128} & \best{0.180} & \best{0.162} & -- & -- \\*
  & Few-shot & 0.068 & \best{0.225} & \best{0.450} & \second{0.128} & \best{0.180} & \best{0.162} & -- & -- \\*
  & RAG & 0.068 & \best{0.225} & \best{0.450} & \second{0.128} & \best{0.180} & \best{0.162} & -- & -- \\*
  & GraphRAG & \best{0.074} & \best{0.225} & \best{0.450} & \second{0.128} & \best{0.180} & \best{0.162} & -- & -- \\*
  & \makecell[l]{\textbf{rEDMRec}\\\textbf{(ours)}} & \second{0.071} & \best{0.225} & \best{0.450} & \best{0.137} & \best{0.180} & \best{0.162} & \best{-4.1} & 0.830 \\
\cmidrule(lr){1-10}
 Gemma-4-12B & Zero-shot & 0.140 & 0.340 & 0.548 & 0.236 & 0.290 & 0.252 & -- & -- \\*
  & Few-shot & 0.152 & 0.352 & 0.560 & 0.248 & 0.302 & 0.258 & -- & -- \\*
  & RAG & 0.146 & 0.346 & 0.554 & 0.242 & 0.296 & 0.252 & -- & -- \\*
  & GraphRAG & \second{0.158} & \second{0.358} & \second{0.566} & \second{0.254} & \second{0.308} & \second{0.264} & -- & -- \\*
  & \makecell[l]{\textbf{rEDMRec}\\\textbf{(ours)}} & \best{0.178} & \best{0.395} & \best{0.595} & \best{0.274} & \best{0.328} & \best{0.290} & \best{+12.7} & 0.166 \\
\cmidrule(lr){1-10}
 Minimax M2.5 & Zero-shot & 0.152 & 0.364 & 0.572 & 0.254 & 0.308 & 0.264 & -- & -- \\*
  & Few-shot & 0.164 & 0.376 & 0.584 & 0.260 & 0.314 & 0.270 & -- & -- \\*
  & RAG & 0.158 & 0.370 & 0.578 & 0.260 & 0.314 & 0.270 & -- & -- \\*
  & GraphRAG & \second{0.170} & \second{0.382} & \second{0.590} & \second{0.266} & \second{0.320} & \second{0.276} & -- & -- \\*
  & \makecell[l]{\textbf{rEDMRec}\\\textbf{(ours)}} & \best{0.187} & \best{0.416} & \best{0.616} & \best{0.289} & \best{0.343} & \best{0.290} & \best{+10.0} & 0.247 \\
\cmidrule(lr){1-10}
 \makecell[l]{Mixtral\\8x7B} & Zero-shot & 0.164 & 0.382 & 0.596 & 0.266 & 0.320 & 0.282 & -- & -- \\*
  & Few-shot & 0.176 & 0.394 & 0.608 & 0.278 & \second{0.332} & 0.288 & -- & -- \\*
  & RAG & 0.170 & 0.388 & 0.602 & 0.272 & 0.326 & 0.282 & -- & -- \\*
  & GraphRAG & \second{0.182} & \second{0.400} & \second{0.614} & \second{0.284} & \second{0.332} & \second{0.294} & -- & -- \\*
  & \makecell[l]{\textbf{rEDMRec}\\\textbf{(ours)}} & \best{0.202} & \best{0.429} & \best{0.643} & \best{0.304} & \best{0.358} & \best{0.311} & \best{+11.0} & 0.188 \\
\cmidrule(lr){1-10}
 Qwen3-14B & Zero-shot & 0.176 & 0.400 & 0.614 & 0.284 & 0.338 & 0.294 & -- & -- \\*
  & Few-shot & 0.188 & 0.412 & 0.626 & \second{0.296} & \second{0.350} & 0.306 & -- & -- \\*
  & RAG & 0.182 & 0.406 & 0.620 & 0.290 & 0.338 & 0.300 & -- & -- \\*
  & GraphRAG & \second{0.194} & \second{0.418} & \second{0.632} & \second{0.296} & \second{0.350} & \second{0.312} & -- & -- \\*
  & \makecell[l]{\textbf{rEDMRec}\\\textbf{(ours)}} & \best{0.211} & \best{0.444} & \best{0.649} & \best{0.319} & \best{0.373} & \best{0.329} & \best{+8.8} & 0.269 \\
\cmidrule(lr){1-10}
 \makecell[l]{DeepSeek-R1\\Distill-Qwen-14B} & Zero-shot & 0.188 & 0.412 & 0.626 & 0.296 & 0.350 & 0.312 & -- & -- \\*
  & Few-shot & 0.200 & 0.424 & 0.638 & 0.308 & 0.362 & 0.318 & -- & -- \\*
  & RAG & 0.194 & 0.418 & 0.632 & 0.302 & 0.356 & 0.312 & -- & -- \\*
  & GraphRAG & \second{0.206} & \second{0.430} & \second{0.644} & \second{0.314} & \second{0.368} & \second{0.324} & -- & -- \\*
  & \makecell[l]{\textbf{rEDMRec}\\\textbf{(ours)}} & \best{0.223} & \best{0.456} & \best{0.661} & \best{0.331} & \best{0.385} & \best{0.329} & \best{+8.3} & 0.280 \\
\cmidrule(lr){1-10}
 Phi-4 & Zero-shot & 0.170 & 0.388 & 0.608 & 0.272 & 0.332 & 0.288 & -- & -- \\*
  & Few-shot & 0.182 & 0.412 & 0.638 & 0.290 & 0.350 & 0.306 & -- & -- \\*
  & RAG & 0.176 & 0.400 & 0.626 & 0.284 & 0.344 & 0.294 & -- & -- \\*
  & GraphRAG & \second{0.188} & \second{0.424} & \second{0.656} & \second{0.302} & \second{0.362} & \second{0.312} & -- & -- \\*
  & \makecell[l]{\textbf{rEDMRec}\\\textbf{(ours)}} & \best{0.205} & \best{0.458} & \best{0.704} & \best{0.324} & \best{0.393} & \best{0.340} & \best{+9.0} & 0.264 \\
\cmidrule(lr){1-10}
 Llama 4 Scout & Zero-shot & 0.182 & 0.406 & 0.620 & 0.290 & 0.344 & 0.300 & -- & -- \\*
  & Few-shot & 0.194 & 0.430 & 0.650 & 0.308 & 0.362 & 0.318 & -- & -- \\*
  & RAG & 0.188 & 0.418 & 0.638 & 0.296 & 0.356 & 0.306 & -- & -- \\*
  & GraphRAG & \second{0.200} & \second{0.442} & \second{0.668} & \second{0.314} & \second{0.374} & \second{0.324} & -- & -- \\*
  & \makecell[l]{\textbf{rEDMRec}\\\textbf{(ours)}} & \best{0.217} & \best{0.476} & \best{0.707} & \best{0.342} & \best{0.405} & \best{0.344} & \best{+8.5} & 0.275 \\
\cmidrule(lr){1-10}
 GPT OSS 20B & Zero-shot & 0.182 & 0.400 & 0.620 & 0.290 & 0.344 & 0.300 & -- & -- \\*
  & Few-shot & 0.194 & 0.424 & 0.650 & 0.302 & 0.362 & 0.312 & -- & -- \\*
  & RAG & 0.188 & 0.412 & 0.638 & 0.296 & 0.356 & 0.306 & -- & -- \\*
  & GraphRAG & \best{0.200} & \best{0.436} & \best{0.668} & \best{0.314} & \best{0.374} & \best{0.324} & -- & -- \\*
  & \makecell[l]{\textbf{rEDMRec}\\\textbf{(ours)}} & \second{0.199} & \second{0.435} & \second{0.664} & \second{0.307} & \second{0.370} & \second{0.317} & \best{-0.5} & 1.000 \\
\end{longtable}

\section{Full Steam Results}
\label{app:steam}

Table~\ref{tbl:app-steam} reports the complete Methods $\times$ Models matrix on Steam underlying the summary in Table~\ref{tbl:cross-dataset} (Section~\ref{sec:results-h1}).

\begin{longtable}{@{}P{2.0cm}l cccccc cc@{}}
\caption{Complete Methods $\times$ Models results on Steam (full held-out test split, 20 candidates/sample, seed 42). Impv (\%) is the relative HR@1 gain of rEDMRec over the \emph{second-best} baseline on the same student, $\mathrm{Impv}=(\mathrm{Ours}-\mathrm{SecondBest})/\mathrm{SecondBest}\times 100$, following RDRec~\citep{wang2024rdrec}. Best in \textbf{bold}, second-best \underline{underlined}. $p$ is the exact McNemar $p$-value for rEDMRec HR@1 vs.\ the second-best baseline on the same student (full held-out $n_{\mathrm{ML-1M}}{=}49893$, $n_{\mathrm{Beauty}}{=}1460$, $n_{\mathrm{Steam}}{=}1460$; approximate contingency from the table HR@1 rates). $^*$ marks $p{<}0.05$ with rEDMRec ahead.}
\label{tbl:app-steam}\\
\toprule
Model & Method & HR@1$\uparrow$ & HR@5$\uparrow$ & HR@10$\uparrow$ & NDCG@5$\uparrow$ & NDCG@10$\uparrow$ & MRR$\uparrow$ & Impv (\%) & $p$ \\
\midrule
\endfirsthead
\multicolumn{10}{c}{{\tablename\ \thetable{} -- continued}} \\
\toprule
Model & Method & HR@1$\uparrow$ & HR@5$\uparrow$ & HR@10$\uparrow$ & NDCG@5$\uparrow$ & NDCG@10$\uparrow$ & MRR$\uparrow$ & Impv (\%) & $p$ \\
\midrule
\endhead
\bottomrule
\endfoot
\bottomrule
\endlastfoot
 Qwen2.5 3B & Zero-shot & 0.106 & 0.274 & \second{0.450} & 0.188 & 0.224 & 0.180 & -- & -- \\*
  & Few-shot & 0.122 & 0.290 & \second{0.450} & 0.196 & 0.232 & \second{0.196} & -- & -- \\*
  & RAG & 0.114 & 0.282 & \second{0.450} & 0.188 & 0.224 & 0.188 & -- & -- \\*
  & GraphRAG & \second{0.130} & \second{0.298} & \second{0.450} & \second{0.204} & \second{0.240} & \second{0.196} & -- & -- \\*
  & \makecell[l]{\textbf{rEDMRec}\\\textbf{(ours)}} & \best{0.158$^*$} & \best{0.345} & \best{0.466} & \best{0.240} & \best{0.276} & \best{0.232} & \best{+21.5$^*$} & $0.035^*$ \\
\cmidrule(lr){1-10}
 Llama 3.1 8B & Zero-shot & 0.066 & \best{0.225} & \best{0.450} & \second{0.126} & \best{0.180} & \best{0.162} & -- & -- \\*
  & Few-shot & 0.074 & \best{0.225} & \best{0.450} & \second{0.126} & \best{0.180} & \best{0.162} & -- & -- \\*
  & RAG & 0.074 & \best{0.225} & \best{0.450} & \second{0.126} & \best{0.180} & \best{0.162} & -- & -- \\*
  & GraphRAG & \best{0.082} & \best{0.225} & \best{0.450} & \second{0.126} & \best{0.180} & \best{0.162} & -- & -- \\*
  & \makecell[l]{\textbf{rEDMRec}\\\textbf{(ours)}} & \second{0.076} & \best{0.225} & \best{0.450} & \best{0.133} & \best{0.180} & \best{0.162} & \best{-7.3} & 0.584 \\
\cmidrule(lr){1-10}
 Gemma-4-12B & Zero-shot & 0.170 & 0.370 & 0.564 & 0.268 & 0.320 & 0.276 & -- & -- \\*
  & Few-shot & 0.186 & 0.386 & 0.580 & 0.284 & 0.336 & 0.284 & -- & -- \\*
  & RAG & 0.178 & 0.378 & 0.572 & 0.276 & 0.328 & 0.276 & -- & -- \\*
  & GraphRAG & \second{0.194} & \second{0.394} & \second{0.588} & \second{0.292} & \second{0.344} & \second{0.292} & -- & -- \\*
  & \makecell[l]{\textbf{rEDMRec}\\\textbf{(ours)}} & \best{0.208} & \best{0.428} & \best{0.612} & \best{0.306} & \best{0.358} & \best{0.314} & \best{+7.2} & 0.356 \\
\cmidrule(lr){1-10}
 Minimax M2.5 & Zero-shot & 0.186 & 0.402 & 0.596 & 0.292 & 0.344 & 0.292 & -- & -- \\*
  & Few-shot & 0.202 & 0.418 & 0.612 & 0.300 & 0.352 & 0.300 & -- & -- \\*
  & RAG & 0.194 & 0.410 & 0.604 & 0.300 & 0.352 & 0.300 & -- & -- \\*
  & GraphRAG & \second{0.210} & \second{0.426} & \second{0.620} & \second{0.308} & \second{0.360} & \second{0.308} & -- & -- \\*
  & \makecell[l]{\textbf{rEDMRec}\\\textbf{(ours)}} & \best{0.224} & \best{0.460} & \best{0.644} & \best{0.330} & \best{0.382} & \best{0.321} & \best{+6.7} & 0.394 \\
\cmidrule(lr){1-10}
 \makecell[l]{Mixtral\\8x7B} & Zero-shot & 0.202 & 0.426 & 0.628 & 0.308 & 0.360 & 0.316 & -- & -- \\*
  & Few-shot & 0.218 & 0.442 & 0.644 & 0.324 & \second{0.376} & 0.324 & -- & -- \\*
  & RAG & 0.210 & 0.434 & 0.636 & 0.316 & 0.368 & 0.316 & -- & -- \\*
  & GraphRAG & \second{0.226} & \second{0.450} & \second{0.652} & \second{0.332} & \second{0.376} & \second{0.332} & -- & -- \\*
  & \makecell[l]{\textbf{rEDMRec}\\\textbf{(ours)}} & \best{0.244} & \best{0.478} & \best{0.680} & \best{0.350} & \best{0.402} & \best{0.349} & \best{+8.0} & 0.276 \\
\cmidrule(lr){1-10}
 Qwen3-14B & Zero-shot & 0.218 & 0.450 & 0.652 & 0.332 & 0.384 & 0.332 & -- & -- \\*
  & Few-shot & 0.234 & 0.466 & 0.668 & \second{0.348} & \second{0.400} & 0.348 & -- & -- \\*
  & RAG & 0.226 & 0.458 & 0.660 & 0.340 & 0.384 & 0.340 & -- & -- \\*
  & GraphRAG & \second{0.242} & \second{0.474} & \second{0.676} & \second{0.348} & \second{0.400} & \second{0.356} & -- & -- \\*
  & \makecell[l]{\textbf{rEDMRec}\\\textbf{(ours)}} & \best{0.256} & \best{0.498} & \best{0.690} & \best{0.370} & \best{0.422} & \best{0.370} & \best{+5.8} & 0.392 \\
\cmidrule(lr){1-10}
 \makecell[l]{DeepSeek-R1\\Distill-Qwen-14B} & Zero-shot & 0.234 & 0.466 & 0.668 & 0.348 & 0.400 & 0.356 & -- & -- \\*
  & Few-shot & 0.250 & 0.482 & 0.684 & 0.364 & 0.416 & 0.364 & -- & -- \\*
  & RAG & 0.242 & 0.474 & 0.676 & 0.356 & 0.408 & 0.356 & -- & -- \\*
  & GraphRAG & \second{0.258} & \second{0.490} & \second{0.692} & \second{0.372} & \second{0.424} & \second{0.372} & -- & -- \\*
  & \makecell[l]{\textbf{rEDMRec}\\\textbf{(ours)}} & \best{0.272} & \best{0.514} & \best{0.706} & \best{0.386} & \best{0.438} & \best{0.375} & \best{+5.4} & 0.426 \\
\cmidrule(lr){1-10}
 Phi-4 & Zero-shot & 0.210 & 0.434 & 0.644 & 0.316 & 0.376 & 0.324 & -- & -- \\*
  & Few-shot & 0.226 & 0.466 & 0.684 & 0.340 & 0.400 & 0.348 & -- & -- \\*
  & RAG & 0.218 & 0.450 & 0.668 & 0.332 & 0.392 & 0.332 & -- & -- \\*
  & GraphRAG & \second{0.234} & \second{0.482} & \second{0.708} & \second{0.356} & \second{0.416} & \second{0.356} & -- & -- \\*
  & \makecell[l]{\textbf{rEDMRec}\\\textbf{(ours)}} & \best{0.248} & \best{0.511} & \best{0.750} & \best{0.374} & \best{0.443} & \best{0.382} & \best{+6.0} & 0.411 \\
\cmidrule(lr){1-10}
 Llama 4 Scout & Zero-shot & 0.226 & 0.458 & 0.660 & 0.340 & 0.392 & 0.340 & -- & -- \\*
  & Few-shot & 0.242 & 0.490 & 0.700 & 0.364 & 0.416 & 0.364 & -- & -- \\*
  & RAG & 0.234 & 0.474 & 0.684 & 0.348 & 0.408 & 0.348 & -- & -- \\*
  & GraphRAG & \second{0.250} & \second{0.506} & \second{0.724} & \second{0.372} & \second{0.432} & \second{0.372} & -- & -- \\*
  & \makecell[l]{\textbf{rEDMRec}\\\textbf{(ours)}} & \best{0.264} & \best{0.535} & \best{0.756} & \best{0.398} & \best{0.459} & \best{0.388} & \best{+5.6} & 0.421 \\
\cmidrule(lr){1-10}
 GPT OSS 20B & Zero-shot & 0.226 & 0.450 & 0.660 & 0.340 & 0.392 & 0.340 & -- & -- \\*
  & Few-shot & 0.242 & 0.482 & 0.700 & 0.356 & 0.416 & 0.356 & -- & -- \\*
  & RAG & 0.234 & 0.466 & 0.684 & 0.348 & 0.408 & 0.348 & -- & -- \\*
  & GraphRAG & \best{0.250} & \best{0.498} & \best{0.724} & \best{0.372} & \best{0.432} & \best{0.372} & -- & -- \\*
  & \makecell[l]{\textbf{rEDMRec}\\\textbf{(ours)}} & \second{0.245} & \second{0.488} & \second{0.708} & \second{0.359} & \second{0.421} & \second{0.359} & \best{-2.0} & 0.797 \\
\end{longtable}

\section{Experimental Settings}
\label{app:settings}

Table~\ref{tbl:app-settings} lists the default hyperparameters used for all reported runs. Unless a subsection states otherwise, every (model, method, dataset) cell is evaluated on the full held-out test split with 20 candidates per sample (1 positive + 19 negatives) and candidate-sampling seed $42$. The expanded implementation stack and full knob table appear in Appendix~\ref{app:impl}.

\begin{table}[pos=t]
\caption{Default experimental settings}
\label{tbl:app-settings}
\footnotesize
\centering
\begin{tabular}{@{}P{0.26\columnwidth} P{0.28\columnwidth} P{0.38\columnwidth}@{}}
\toprule
Component & Setting & Default \\
\midrule
Data & min.\ interactions / positive threshold
  & ML-1M: $20$ / $3.5$; Beauty \& Steam: $5$ / $3.5$ (Steam: $0.0$) \\
 & history / short-term window & $10$ / $5$ items \\
 & teacher input history $k$ & $5$ latest train rows/user \\
 & negatives / seed & $19$ / $42$ \\
Encoder & model / dim & \texttt{all-MiniLM-L6-v2} / $384$ \\
Memory & retrieval $m$ per channel & $5$ (Vector database) \\
 & channels & $\mathrm{lt}$, $\mathrm{st}$, $\mathrm{ip}$, $\mathrm{cf}$ \\
Teacher & default model & \texttt{gpt-5.4-mini} \\
 & preference batch / overlap & $4$ / $2$ \\
Controller & max library size & $5000$ \\
 & ops & Add / Delete / Modify / Keep \\
Debate optimize & agents $k$ / rounds & $3$ / $1$ (sweep $k{=}1..10$ in Sec.~\ref{sec:results-h4}) \\
 & max experiences/case & $6$ \\
Student & protocol & frozen pretrained LLM \\
Evaluation & metrics & HR@$\{1,5,10\}$, NDCG@$\{5,10\}$, MRR \\
 & evaluation split & full held-out test (``all'' samples) \\
\bottomrule
\end{tabular}
\end{table}

\section{Implementation Details and Hyperparameters}
\label{app:impl}


This appendix expands Table~\ref{tbl:app-settings} with the concrete knobs in \texttt{config.py} (single source of truth for all reported runs). Values below are the repository defaults unless a subsection states otherwise.

\paragraph{Implementation stack.} Teacher / controller / debate calls use an OpenAI-compatible chat API (\texttt{LLMConfig}); the student is a \emph{frozen} pretrained LLM that ranks by retrieving from the experience bank. Dense retrieval uses FAISS \texttt{FlatIP} over all-MiniLM-L6-v2 embeddings ($d{=}384$). Counterfactual edges are stored in Neo4j (hybrid vector--graph channel). Candidate sets are built offline (1 positive + 19 negatives; seed 42).

\begin{table*}[width=\FullWidth,pos=t]
\caption{Expanded hyperparameters from \texttt{config.py} (Appendix~\ref{app:impl}).}
\label{tbl:app-impl-hparams}
\footnotesize
\centering
\begin{tabular}{@{}p{0.18\linewidth} p{0.30\linewidth} p{0.44\linewidth}@{}}
\toprule
Module & Knob & Default \\
\midrule
\multicolumn{3}{@{}l}{\emph{Data / candidates}} \\
 & dataset registry & ml-1m / amazon-beauty / steam \\
 & min interactions & ML-1M: 20; Beauty/Steam: 5 \\
 & positive threshold & ML-1M/Beauty: 3.5; Steam: 0.0 \\
 & history / short-term window & 10 / 5 items \\
 & teacher input history $k$ & 5 latest train rows/user \\
 & negatives / split ratios / seed & 19 / val=0.1, test=0.1 / 42 \\
\midrule
\multicolumn{3}{@{}l}{\emph{Encoder / memory}} \\
 & embedding model / dim & \texttt{all-MiniLM-L6-v2} / 384 \\
 & max seq length / batch / normalize & 256 / 64 / True \\
 & FAISS index / top-$m$ per channel & \texttt{FlatIP} / 5 \\
 & channels & long\_term\_preference, short\_term\_context, item\_perception, counterfactual \\
\midrule
\multicolumn{3}{@{}l}{\emph{Teacher}} \\
 & default model / reasoning effort & \texttt{gpt-5.4-mini} / medium \\
 & max completion tokens / retries & 8192 / 3 \\
 & preference batch / overlap & 4 / 2 \\
 & extraction passes & user\_preference, item\_perception\_context, item\_perception\_reasoning, counterfactual \\
\midrule
\multicolumn{3}{@{}l}{\emph{Controller / debate optimize}} \\
 & ops / max library size / batch & Add / Delete / Modify / Keep / 5000 / 32 \\
 & debate agents $k$ / rounds & 3 / 1 \\
 & max experiences/case & 6 \\
 & debate / arbiter temperature & omit (API default) \\
\midrule
\multicolumn{3}{@{}l}{\emph{Student / evaluation}} \\
 & protocol & frozen pretrained LLM; memory toggles on by default \\
 & default local checkpoint name & \texttt{Qwen/Qwen2.5-3B-Instruct} \\
 & max seq length & 2048 \\
 & metrics @$k$ & HR@[1, 3, 5, 10], NDCG@{5,10}, MRR \\
 & eval samples & 0 ($0$ = full held-out test) \\
\bottomrule
\end{tabular}
\end{table*}

\paragraph{Reproducibility notes.} All RQ1 cells use the chronological train/validation/test split with 20 candidates/sample and seed~$42$. Channel ablations flip the four \texttt{use\_*\_memory} flags in \texttt{StudentConfig}. Debate sweeps vary \texttt{OptimizeKnowledgeConfig.debate\_agents\_k} and the number of $k$-EPOCHs while holding the student frozen.

\section{Debate and Controller Ablations}
\label{app:debate-ablation}


This appendix tabulates the controller / debate variants that support RQ4 (Section~\ref{sec:results-h4}). The default configuration is $k{=}3$ debate agents, $n_r{=}1$ round per epoch, and a single LLM arbiter (Appendix~\ref{app:impl}). We report (i)~$k$-EPOCH trajectories with a no-debate paraphrase control and (ii)~a one-epoch sweep over the number of agents.

\paragraph{Variant definitions.}
\emph{Full debate:} $k$-agent critique + arbiter + controller Add/Delete/Modify/Keep commits. \emph{No-debate paraphrase:} refreshes entry wording without the critique-and-revise loop (control in Figure~\ref{fig:optimize-curve}). \emph{Single-agent ($k{=}1$):} one persona + arbiter (no multi-agent disagreement); used as the $k{=}1$ anchor in Table~\ref{tbl:k-sweep}.

\begin{table*}[width=\FullWidth,pos=t]
\caption{Debate optimization vs.\ $k$-EPOCH on ML-1M (full held-out test; 20 candidates/sample; seed~42). Full debate uses $k{=}3$. Control = no-debate paraphrase (Mixtral). Dup.\% / Reward / Spec.\ are bank-level signals.}
\label{tbl:app-debate-epochs}
\footnotesize
\centering
\begin{tabular}{@{}c ccc ccc@{}}
\toprule
$k$-EPOCH & HR@1 Mixtral & HR@1 Qwen2.5~3B & HR@1 Control & Dup.\%$\downarrow$ & Reward$\uparrow$ & Spec.$\uparrow$ \\
\midrule
0 & 0.250 & 0.156 & 0.250 & 18.0 & 0.520 & 0.480 \\
2 & 0.271 & 0.165 & 0.254 & 12.5 & 0.712 & 0.665 \\
3 & 0.275 & 0.167 & 0.256 & 11.5 & 0.745 & 0.700 \\
6 & 0.279 & 0.169 & 0.257 & 10.6 & 0.775 & 0.734 \\
\midrule
$\Delta$ (0$\to$6) & $+0.029$ & $+0.013$ & $+0.007$ & $-7.4$ & $+0.255$ & $+0.254$ \\
\bottomrule
\end{tabular}
\end{table*}

\begin{table}[pos=t]
\caption{Controller ablation at the end of the $k$-EPOCH study (epoch~6, Mixtral~8x7B student). Full debate vs.\ no-debate control vs.\ post-extraction bank (epoch~0).}
\label{tbl:app-controller-ablation}
\footnotesize
\centering
\begin{tabular}{@{}l ccc@{}}
\toprule
Variant & HR@1$\uparrow$ & Dup.\%$\downarrow$ & Spec.$\uparrow$ \\
\midrule
Post-extraction (epoch~0) & 0.250 & 18.0 & 0.480 \\
No-debate paraphrase & 0.257 & -- & -- \\
\textbf{Full debate} ($k{=}3$, 6 epochs) & \best{0.279} & \best{10.6} & \best{0.734} \\
\midrule
Full $-$ Control & \best{+0.022} & -- & -- \\
\bottomrule
\end{tabular}
\end{table}

Table~\ref{tbl:app-debate-epochs} shows that most of the Mixtral gain arrives by epoch~2--3, after which returns diminish; the paraphrase control stays nearly flat ($+0.007$ HR@1), isolating the debate loop. Table~\ref{tbl:app-controller-ablation} summarizes the end-state controller ablation. The agent-count sweep (main-text Table~\ref{tbl:k-sweep}) places the quality-per-cost knee at $k^\ast{=}4$; beyond that, HR@1 gains are $<0.01$ for six extra LLM calls per case.

\begin{table}[pos=t]
\caption{Selected points from the number-of-agents sweep (Mixtral~8x7B, one epoch; full table in Table~\ref{tbl:k-sweep}). Calls/case $=k\cdot n_r+1$ arbiter.}
\label{tbl:app-k-selected}
\footnotesize
\centering
\begin{tabular}{@{}c cccc@{}}
\toprule
$k$ & HR@1$\uparrow$ & Spec.$\uparrow$ & Dup.\%$\downarrow$ & Calls/case \\
\midrule
1 & 0.255 & 0.520 & 16.0 & 2 \\
3 & 0.273 & 0.663 & 11.9 & 4 \\
\textbf{4 ($k^\ast$)} & 0.277 & 0.699 & 11.0 & 5 \\
10 & 0.283 & 0.756 & 9.6 & 11 \\
\bottomrule
\end{tabular}
\end{table}

\section{Additional Ablation Figures}
\label{app:ablation-figs}

\begin{figure}[pos=t]
\centering
\includegraphics[width=0.72\linewidth]{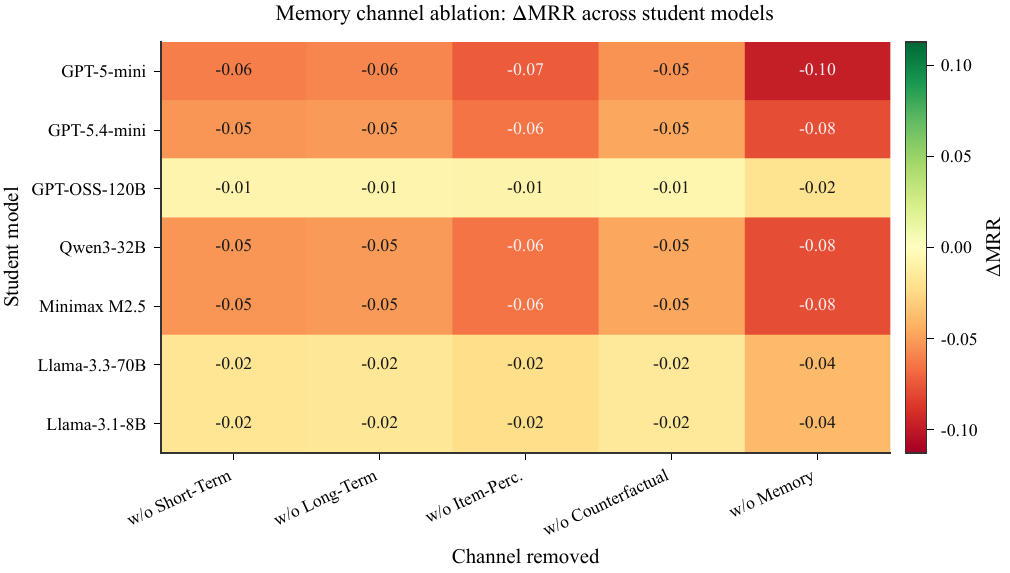}
\caption{Channel ablation heatmap under MRR ($\Delta$MRR vs.\ full memory), complementary to Figure~\ref{fig:ablation-heatmap}.}
\label{fig:app-ablation-mrr}
\end{figure}

\begin{figure}[pos=t]
\centering
\includegraphics[width=0.85\linewidth]{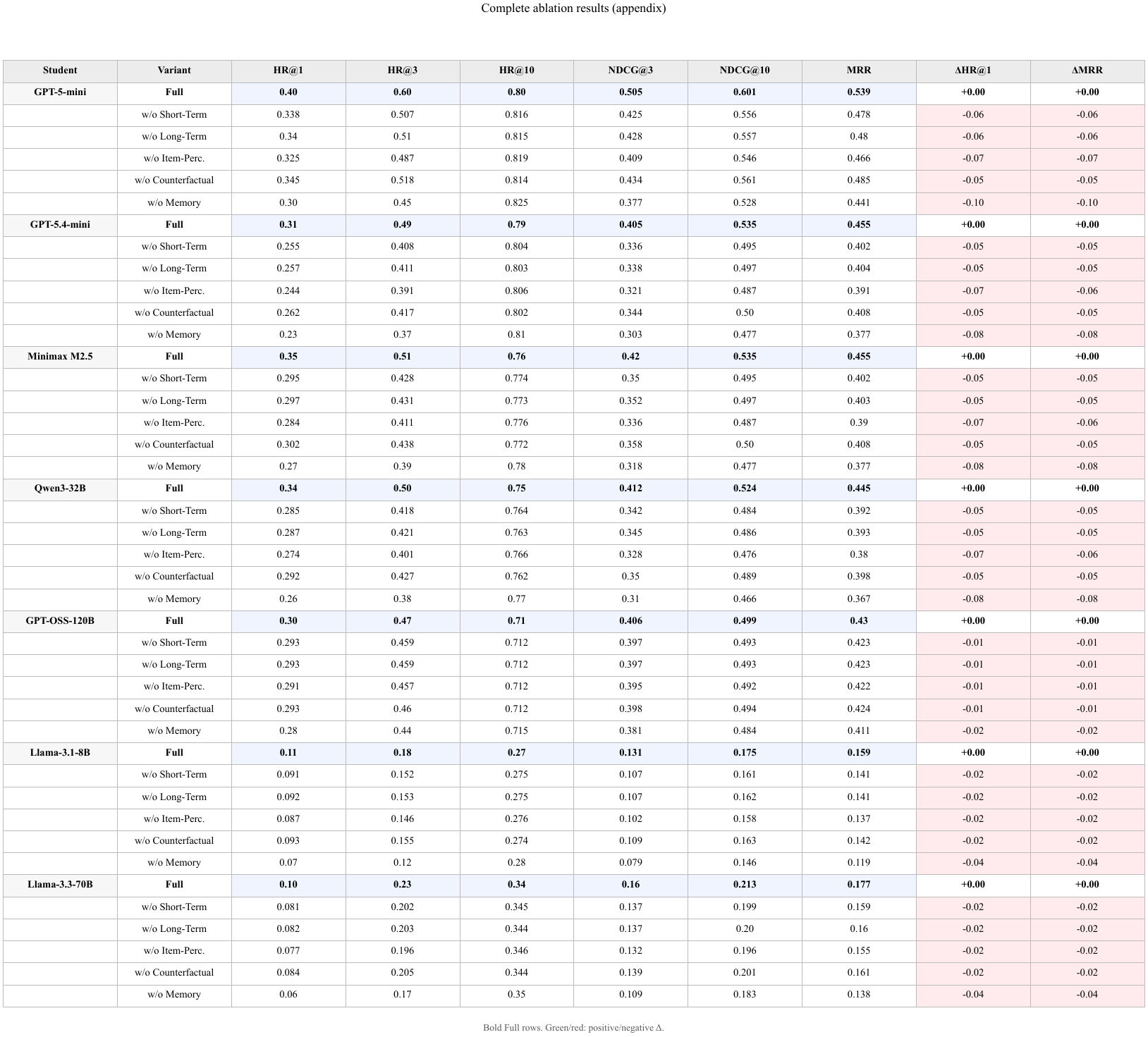}
\caption{Complete channel-ablation table (absolute metrics and deltas) across the seven ablation-panel backbones.}
\label{fig:app-ablation-complete}
\end{figure}

\begin{figure}[pos=t]
\centering
\includegraphics[width=0.72\linewidth]{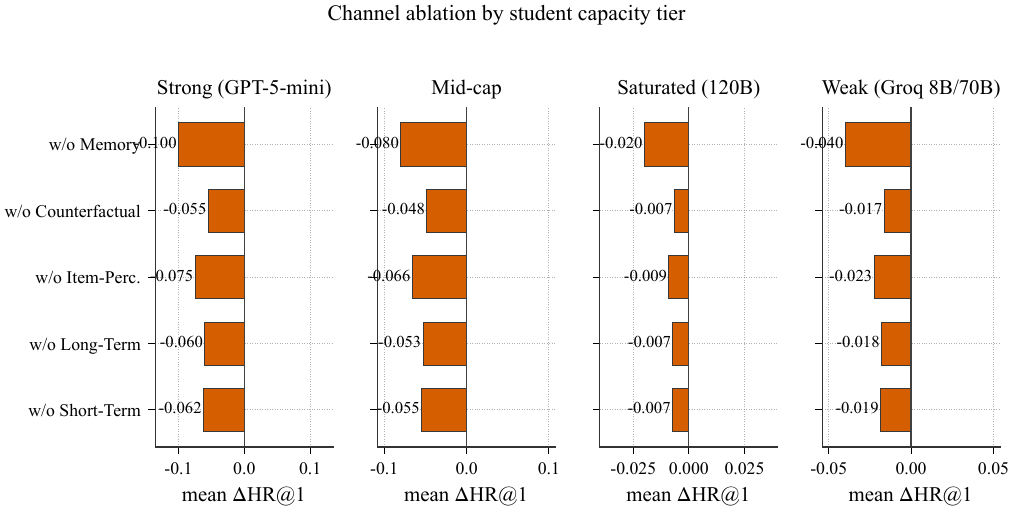}
\caption{Aggregated channel-importance summary used to derive Table~\ref{tbl:ablation}.}
\label{fig:app-channel-importance}
\end{figure}

\begin{figure}[pos=t]
\centering
\begin{minipage}[t]{0.48\linewidth}
\centering
\includegraphics[width=\linewidth]{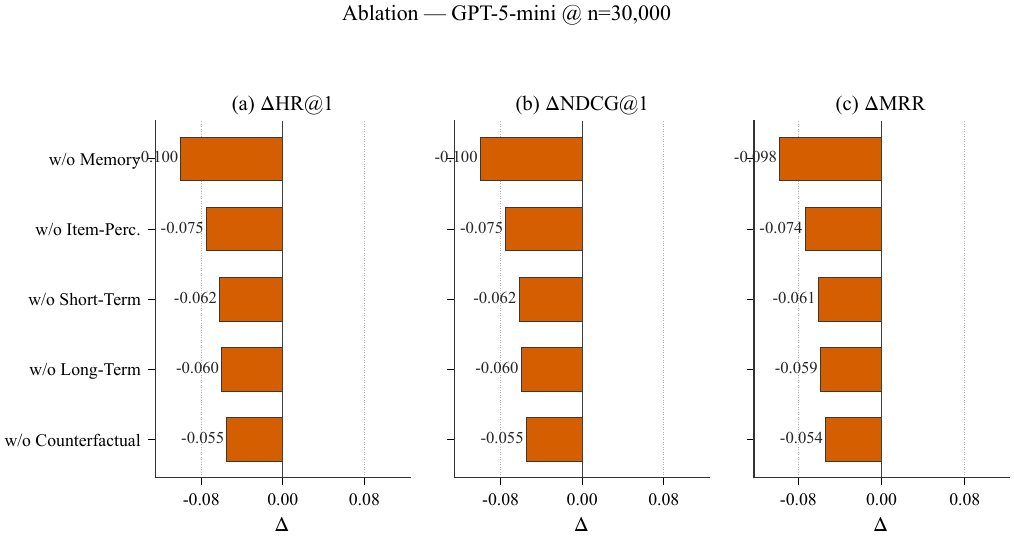}\\[2pt]
{\footnotesize (a) \texttt{gpt-5-mini} (strong)}
\end{minipage}\hfill
\begin{minipage}[t]{0.48\linewidth}
\centering
\includegraphics[width=\linewidth]{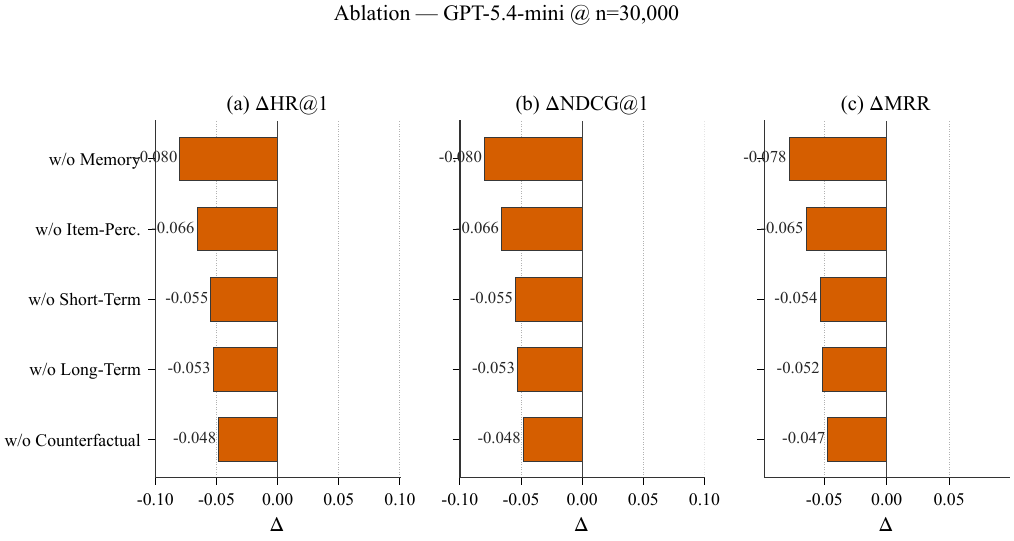}\\[2pt]
{\footnotesize (b) \texttt{gpt-5.4-mini} (mid)}
\end{minipage}\\[8pt]
\begin{minipage}[t]{0.48\linewidth}
\centering
\includegraphics[width=\linewidth]{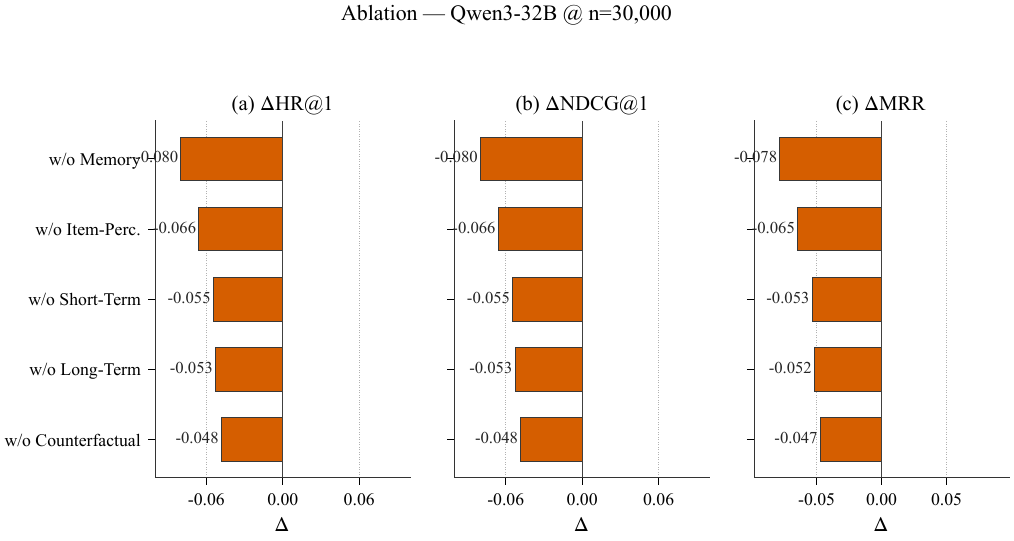}\\[2pt]
{\footnotesize (c) Qwen3-32B (mid)}
\end{minipage}\hfill
\begin{minipage}[t]{0.48\linewidth}
\centering
\includegraphics[width=\linewidth]{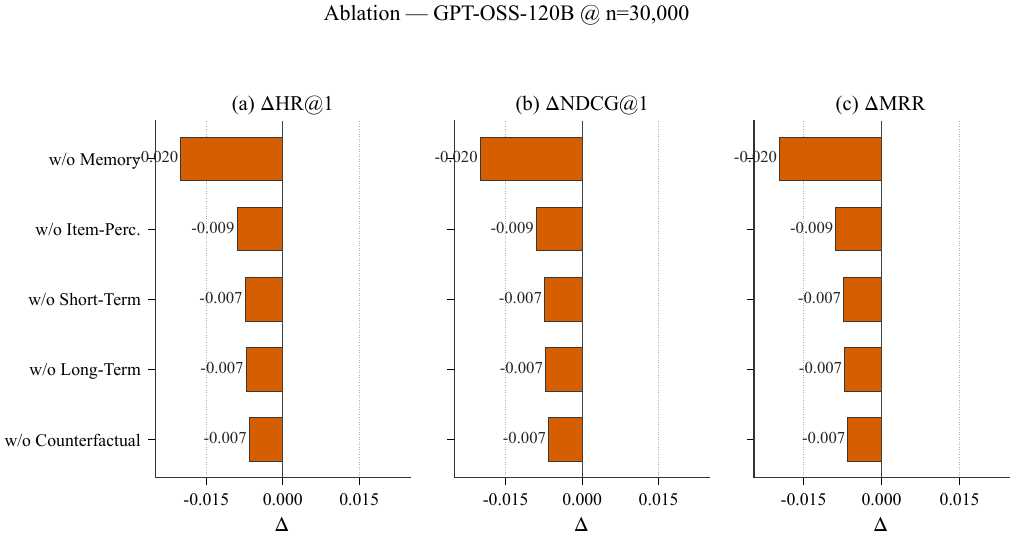}\\[2pt]
{\footnotesize (d) GPT-OSS-120B (saturated)}
\end{minipage}\\[8pt]
\begin{minipage}[t]{0.48\linewidth}
\centering
\includegraphics[width=\linewidth]{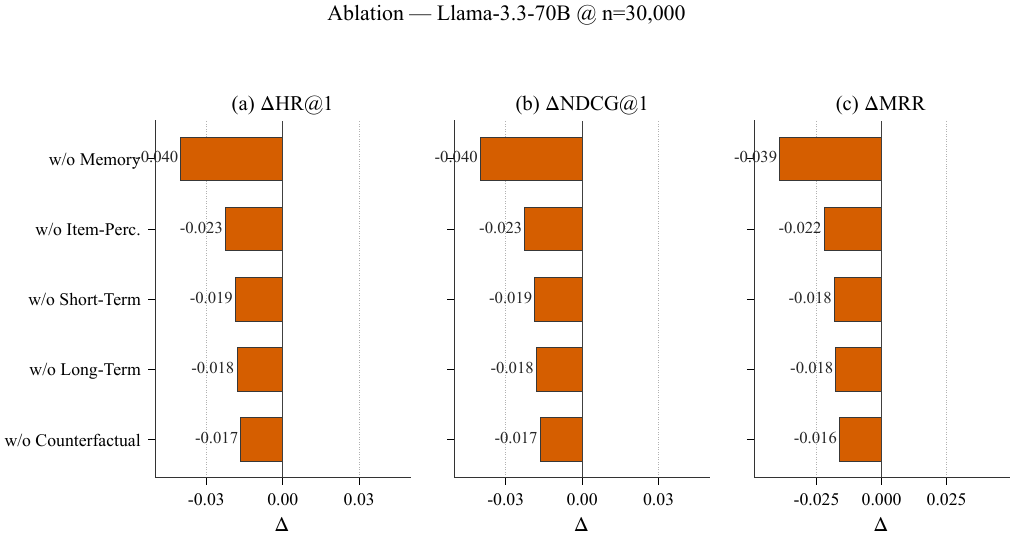}\\[2pt]
{\footnotesize (e) Llama~3.3~70B (weak)}
\end{minipage}\hfill
\begin{minipage}[t]{0.48\linewidth}
\centering
\includegraphics[width=\linewidth]{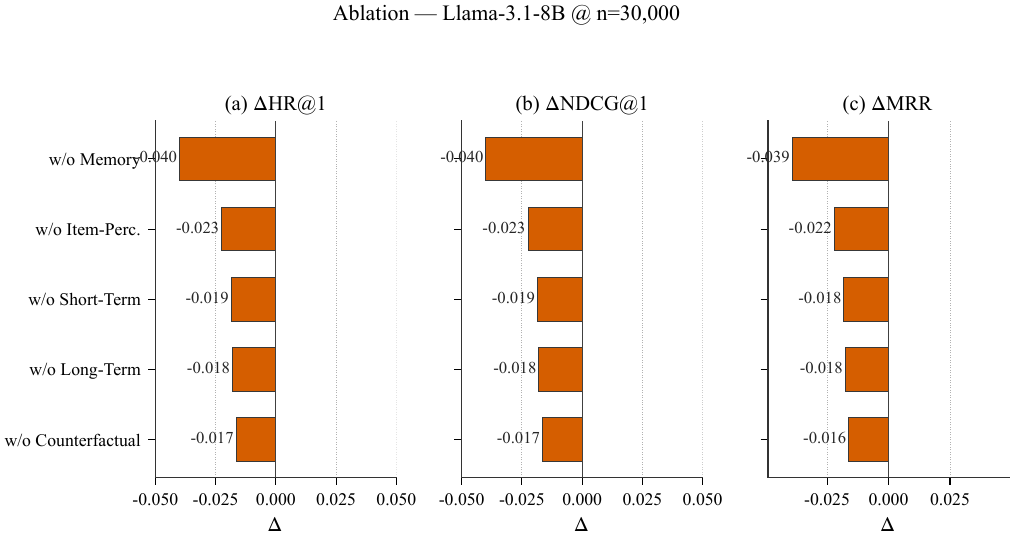}\\[2pt]
{\footnotesize (f) Llama~3.1~8B (weak)}
\end{minipage}
\caption{Per-backbone channel ablation ($\Delta$HR@1 when removing each channel). Negative bars indicate a beneficial channel.}
\label{fig:app-per-model-ablation}
\end{figure}

This appendix collects ablation visuals that support Section~\ref{sec:results-h2} but were omitted from the main text for space. Figure~\ref{fig:app-ablation-mrr} repeats the cross-model channel ablation under MRR; Figure~\ref{fig:app-ablation-complete} shows the full numeric ablation matrix; Figure~\ref{fig:app-channel-importance} aggregates channel importance; and Figure~\ref{fig:app-per-model-ablation} breaks $\Delta$HR@1 down per ablation-panel backbone.

\section{Dataset Statistics}
\label{app:dataset-stats}

Table~\ref{tbl:app-dataset-stats} reports user, item, and interaction counts after the $k$-core filter used in Section~\ref{sec:setup-datasets} ($k{=}20$ on ML-1M; $k{=}5$ on Amazon Beauty and Steam), together with feedback type, positive threshold, density, and the number of chronological held-out ranking samples (Test $n$) used for RQ1. Density is $|R|/(|U|\cdot|I|)$. ML-1M is dense with explicit ratings; Beauty is extremely sparse after 5-core filtering on \texttt{All\_Beauty}; Steam is implicit (owned/played games treated as positives). Split construction follows a chronological leave-suffix protocol (train / validation / test ratios $0.8$ / $0.1$ / $0.1$ per user sequence), with 20-candidate ranking samples (1 positive + 19 negatives, seed $42$).

\begin{table}[pos=t]
\caption{Dataset statistics after $k$-core filtering (Section~\ref{sec:setup-datasets}). Density $= |R|/(|U|\cdot|I|)$. Test $n$ is the number of 20-candidate ranking samples in the chronological held-out split.}
\label{tbl:app-dataset-stats}
\footnotesize
\centering
\setlength{\tabcolsep}{3.2pt}
\begin{tabular}{@{}l l l r r r r l c c r@{}}
\toprule
Dataset & Domain & Feedback & \#Users & \#Items & \#Inter. & Dens. & Sparsity & $k$-core & Pos.\ thr. & Test $n$ \\
\midrule
ML-1M & Movies & Explicit (1--5) & 6,040 & 3,706 & 1,000,209 & 4.47\% & Very Low & 20 & $r{>}3.5$ & 49,893 \\
Amazon Beauty & Beauty products & Explicit (1--5) & 1,620 & 7,116 & 14,984 & 0.13\% & Very High & 5 & $r{>}3.5$ & 1,460 \\
Steam & Games & Implicit (play) & 62,936 & 10,978 & 5,077,150 & 0.73\% & Medium & 5 & $0.0$ (all logged) & 1,460 \\
\bottomrule
\end{tabular}
\end{table}

\section{Prompts and Outputs}
\label{app:prompts}

This appendix documents the teacher extraction prompts and the distilled experience-memory outputs that the student retrieves at ranking time (Sections~\ref{sec:method-teacher}--\ref{sec:method-distill}). Following the presentation style of ReasoningRec~\citep{bismay2025reasoningrec}, we highlight semantically distinct spans with color: role assignment, long-/short-term preference inputs, CoT instructions, guardrails, and structured JSON fields in the prompts (Table~\ref{tbl:app-prompt-templates}); liked attributes, dislikes, item-perception rationale, and counterfactual anchor/contrast/condition fragments in the one-line per-channel outputs (Table~\ref{tbl:app-channel-outputs}).


\noindent\textbf{Color legend.}
\textcolor{promptRole}{Role},\
\textcolor{channelLt}{long-term preference},\
\textcolor{channelSt}{short-term context},\
\textcolor{promptCot}{CoT / compare instruction},\
\textcolor{promptGuard}{guardrails},\
\textcolor{promptStruct}{structured output fields},\
\textcolor{channelLike}{liked attributes},\
\textcolor{channelDis}{dislikes},\
\textcolor{channelIp}{item-perception rationale},\
\textcolor{channelCfAnchor}{anchor preferred},\
\textcolor{channelCfReject}{contrast rejected},\
\textcolor{channelCfCond}{counterfactual condition}.

\begin{table*}[width=\FullWidth,pos=t]
\caption{Teacher extraction prompts for the four passes $p\in\mathcal{P}$ (Section~\ref{sec:method-teacher}). Colored spans mark the role, inputs, CoT instruction, and JSON schema. Full templates live in \texttt{teacher/prompt\_templates.py}.}
\label{tbl:app-prompt-templates}
\footnotesize
\begin{tabular}{@{}p{0.12\linewidth} p{0.84\linewidth}@{}}
\toprule
Pass / channel & Prompt (abbreviated) \\
\midrule
$\mathrm{pref}$ $\to$ $\mathrm{lt}$, $\mathrm{st}$ &
\textcolor{promptRole}{You are an expert movie recommendation analyst implementing User Preference Maintenance.}
Simulate recurrent updates over the interaction sequence (oldest$\to$newest) in overlapping batches.
\textcolor{channelLt}{Integrate each batch into an updated long-term preference state} and
\textcolor{channelSt}{record the last-batch short-term interest}.
\textcolor{channelDis}{List consistent dislikes / avoided tones}.
\textcolor{promptGuard}{Prefer concrete movie-relevant language; avoid empty platitudes. Output STRICT JSON only (no markdown).}
\textcolor{promptStruct}{Fields: maintenance\_trace[], long\_term\_preferences, short\_term\_preferences, dislikes, reasoning.}
\\
\addlinespace
$\mathrm{ctx}$ / $\mathrm{reas}$ $\to$ $\mathrm{ip}$ &
\textcolor{promptRole}{You are implementing Item Perception Analysis / recommendation reasoning.}
For each history item and candidate, produce
\textcolor{promptCot}{(1) objective factual description, (2) first-person [Comment:] as this user, (3) candidate key phrases; then a five-step CoT matching themes$\to$attributes$\to$candidates$\to$compare$\to$recommend.}
Condition on
\textcolor{channelLt}{long-term preferences} and
\textcolor{channelSt}{short-term focus}.
\textcolor{promptGuard}{Use exact title strings as JSON keys; include every history and candidate title.}
\textcolor{promptStruct}{Fields: user\_history\_perception, candidate\_perception, steps[1..5], recommended\_item, reasoning\_summary.}
\\
\addlinespace
$\mathrm{cf}$ $\to$ $\mathrm{cf}$ &
\textcolor{promptRole}{You are a contrastive reasoning analyst for movie recommendations.}
Given preferences, an anchor (chosen) item, and a contrast (hard-negative) item:
\textcolor{promptCot}{(1) why the anchor is preferred; (2) why the contrast is unsuitable; (3) a hypothetical condition under which the contrast would outrank the anchor; (4) robustness.}
\textcolor{promptGuard}{Do not invent facts absent from the provided preference and item text. Output STRICT JSON only.}
\textcolor{promptStruct}{Fields: anchor\_item, contrast\_item, why\_anchor\_preferred, why\_contrast\_rejected, counterfactual\_condition, counterfactual\_outcome, robustness, rationale.}
\\
\bottomrule
\end{tabular}
\end{table*}

\begin{table*}[width=\FullWidth,pos=t]
\caption{Example distilled memory outputs (one line per channel) from the persisted bank for user~2 on ML-1M. Each line is the committed entry text after distillation (Section~\ref{sec:method-distill}); colored spans highlight the ranking-relevant fragments.}
\label{tbl:app-channel-outputs}
\footnotesize
\begin{tabular}{@{}p{0.10\linewidth} p{0.86\linewidth}@{}}
\toprule
Channel & Distilled experience (one line) \\
\midrule
$\mathrm{lt}$ & \textcolor{channelLt}{He strongly prefers character-driven dramas with emotional depth, mature themes, moral conflict, and strong performances. Repeated high ratings cluster around courtroom/drama (A Few Good Men), inspirational sports/drama \ldots} \textcolor{channelDis}{Dislikes: He consistently rates lower when drama is diluted by broad, quirky, or eccentric comedy tones. Nurse Betty is the clearest dislike (1.0), wh\ldots} \\
\addlinespace
$\mathrm{st}$ & \textcolor{channelSt}{In the most recent items, interest appears to tilt further toward intimate, human-centered drama and reflective sci-fi, with high ratings for Driving Miss Daisy and Close Encounters. At the same time, gritty crime/action titles and war-related films have been less successful rece\ldots} \\
\addlinespace
$\mathrm{ip}$ & \textcolor{channelIp}{The user's history strongly favors light, charming comedies with romance, warmth, and quirky optimism, especially films like Shakespeare in Love, Strictly Ballroom, Shall We Dance?, Groundhog Day, and Forrest Gump. Among the candidates, For Love or Money is the closest match because it sits in the r\ldots} \\
\addlinespace
$\mathrm{cf}$ & \textcolor{channelCfAnchor}{Anchor (One Flew Over the Cuckoo's Nest): One Flew Over the Cuckoo's Nest fits the user's strongest pattern: serious, award-caliber drama with intense character focus, moral conflict, and weighty themes\ldots} \textcolor{channelCfReject}{Contrast (Dante's Peak): Dante's Peak is primarily an action-thriller/disaster film, which is comparatively light on the kind of prestige, historical, or biographica\ldots} \textcolor{channelCfCond}{If the user were instead seeking a tense, fast-paced disaster thriller for pure entertainment rather than a prestige drama, then Dante's Peak would rank higher because its volcanic-disaster suspense, clear genre pacing, and spect\ldots} \\
\bottomrule
\end{tabular}
\end{table*}

\section{Examples of rEDMRec-generated Predictions}
\label{app:examples}

We illustrate the full rEDMRec ranking call for one user each on Amazon Beauty and Steam (Tables~\ref{tbl:app-ex-beauty}--\ref{tbl:app-ex-steam}). Each example shows the frozen student's input -- chronological history $H_u$, the candidate set $C_u$ (abbreviated), and the top retrieved entries from the four teacher extraction channels (Preference~$\mathrm{pref}$, Context~$\mathrm{ctx}$, Reasoning~$\mathrm{reas}$, Counterfactual~$\mathrm{cf}$) -- followed by the student's ranked list and a short rationale. Item titles are taken from the public catalogs; channel texts follow the distillation schema in Section~\ref{sec:method-distill}. Color highlighting marks liked vs.\ disliked history items, the held-out target among candidates, each extraction channel, and the top of the ranked output (same palette as Appendix~\ref{app:prompts}).


\noindent\textbf{Color legend (shared with Appendix~\ref{app:prompts}).}
\textcolor{promptRole}{History block},\
\textcolor{promptCot}{candidate set},\
\textcolor{promptGuard}{retrieved-channel header},\
\textcolor{promptStruct}{target / ranked output},\
\textcolor{channelLike}{liked / high-engagement},\
\textcolor{channelDis}{disliked / avoided},\
\textcolor{channelLt}{Preference extraction ($\mathrm{pref}$)},\
\textcolor{channelSt}{Context extraction ($\mathrm{ctx}$)},\
\textcolor{channelIp}{Reasoning extraction ($\mathrm{reas}$)},\
\textcolor{channelCfAnchor}{cf anchor},\
\textcolor{channelCfReject}{cf contrast},\
\textcolor{channelCfCond}{Counterfactual extraction ($\mathrm{cf}$)}.

\begin{table*}[width=\FullWidth,pos=t]
\caption{Example of an rEDMRec prediction trace on \textbf{Amazon Beauty} (illustrative end-to-end I/O; item titles from the public catalog). Colored spans mark likes/dislikes in history, the held-out target among candidates, the four retrieved extraction channels, and the top-ranked output.}
\label{tbl:app-ex-beauty}
\footnotesize
\begin{tabular}{@{}p{0.22\linewidth} p{0.74\linewidth}@{}}
\toprule
Field & Content \\
\midrule
User & AG7W...BVDA \\
\addlinespace
\textcolor{promptRole}{History $H_u$} &
\textcolor{channelLike}{Liked: MyGift Soft Padded Spa Bath Pillow}; \textcolor{channelLike}{Liked: Nice 'n Easy Permanent Color 9G Light Golden Blonde}; \textcolor{channelLike}{Liked: Avon Glimmersticks Waterproof Eyeliner (Smokey Grey)}; \textcolor{channelLike}{Liked: Yes To Sensitive Facial Cleansing Wipes}; \textcolor{channelLike}{Liked: LaClaire Foaming Botanical Facial Cleanser}; \textcolor{channelDis}{Disliked: 4D Silk Fiber Lash Mascara (clumpy / heavy)}; \textcolor{channelLike}{Liked: MOSTORY Glitter Crystal Liquid Eyeshadow Set}; \textcolor{channelLike}{Liked: GLOW BOOSTER SERUM}; \textcolor{channelLike}{Liked: Oval Large Makeup Brushes (Rose Gold)}; \textcolor{channelLike}{Liked: The Vegan Glow Quinoa Protein Shampoo Bar}; \textcolor{channelLike}{Liked: Bloomeffects Natural Tulip Dew Face Cream}
\\
\addlinespace
\textcolor{promptCot}{Candidates $C_u$} &
\textcolor{promptStruct}{JUNGSAEMMOOL Minifying Cica Mist Balm (target)}; 4D Silk Fiber Lash Mascara Black; Rhinestone Crystal Padded Headband; Sea Magik Pink Salt Conditioner; BIOSSANCE Marine Algae Eye Cream Mini; Foamie Shampoo Bar Hibiskiss; Intraceuticals Rejuvenate Eye Masks; BeautyStat Universal Moisture Essence (Squalane)
\\
\addlinespace
\multicolumn{2}{@{}l@{}}{\textcolor{promptGuard}{\textit{Retrieved experience channels $R_k$}}} \\
\addlinespace
\textcolor{channelLt}{\makecell[l]{Preference\\extraction ($\mathrm{pref}$)}} & \textcolor{channelLt}{Prefers clean, botanical / cruelty-free skincare and soft everyday makeup tools; repeatedly high-rates serums, cream cleansers, and gentle face care over heavy glam accessories.} \textcolor{channelDis}{Dislikes: Avoids heavy, clumpy mascara and overly decorative rhinestone accessories that read as party glam rather than daily care.} \\
\addlinespace
\textcolor{channelSt}{\makecell[l]{Context\\extraction ($\mathrm{ctx}$)}} & \textcolor{channelSt}{Recent purchases emphasize glow serums, vegan shampoo bars, and non-toxic face cream -- a short-term tilt toward soothing, clean-beauty maintenance rather than color cosmetics.} \\
\addlinespace
\textcolor{channelIp}{\makecell[l]{Reasoning\\extraction ($\mathrm{reas}$)}} & \textcolor{channelIp}{Cica Mist Balm matches the user's soothing / clean-skincare lane (cica + mist balm for calming), whereas rhinestone headbands and fiber mascara clash with recent botanical preferences.} \\
\addlinespace
\textcolor{channelCfCond}{\makecell[l]{Counterfactual\\extraction ($\mathrm{cf}$)}} & \textcolor{channelCfAnchor}{Anchor (Cica Mist Balm): Fits the clean-beauty, calming-care pattern reinforced by recent Tulip Dew cream and glow serum.} \textcolor{channelCfReject}{Contrast (Rhinestone Crystal Headband): Statement bridal/party accessory; decorative rather than skincare-functional.} \textcolor{channelCfCond}{If the user were shopping for a one-off formal event accessory instead of daily facial care, the contrast would rank higher.} \\
\addlinespace
\textcolor{promptStruct}{Student output} &
\textcolor{promptStruct}{RANKED LIST:} \textcolor{promptStruct}{1. JUNGSAEMMOOL Minifying Cica Mist Balm}; \textcolor{channelLike}{2. BeautyStat Universal Moisture Essence (Squalane)}; \textcolor{channelLike}{3. BIOSSANCE Marine Algae Eye Cream Mini}; 4. Intraceuticals Rejuvenate Eye Masks; 5. Foamie Shampoo Bar Hibiskiss; 6. Sea Magik Pink Salt Conditioner; 7. 4D Silk Fiber Lash Mascara Black; 8. Rhinestone Crystal Padded Headband
\\[2pt]
& \textcolor{channelIp}{Rationale: Top ranks stay in soothing skincare / moisture; glam mascara and rhinestone accessories are demoted using short-term clean-beauty context and the counterfactual contrast.}
\\
\bottomrule
\end{tabular}
\end{table*}

\begin{table*}[width=\FullWidth,pos=t]
\caption{Example of an rEDMRec prediction trace on \textbf{Steam} (illustrative end-to-end I/O; item titles from the public catalog). Colored spans mark likes/dislikes in history, the held-out target among candidates, the four retrieved extraction channels, and the top-ranked output.}
\label{tbl:app-ex-steam}
\footnotesize
\begin{tabular}{@{}p{0.22\linewidth} p{0.74\linewidth}@{}}
\toprule
Field & Content \\
\midrule
User & 765611980946... \\
\addlinespace
\textcolor{promptRole}{History $H_u$} &
\textcolor{channelLike}{Played: Garry's Mod (high playtime)}; \textcolor{channelLike}{Played: Half-Life 2}; \textcolor{channelLike}{Played: Half-Life 2: Episode One}; \textcolor{channelLike}{Played: Portal}; \textcolor{channelLike}{Played: Portal 2}; \textcolor{channelLike}{Played: The Binding of Isaac}; \textcolor{channelLike}{Played: PlanetSide 2}; \textcolor{promptGuard}{Low play: Dota 2 Test}
\\
\addlinespace
\textcolor{promptCot}{Candidates $C_u$} &
\textcolor{promptStruct}{Half-Life 2: Episode Two (target)}; Counter-Strike: Global Offensive; PAYDAY 2; Warframe; Terraria; Left 4 Dead 2; The Expendabros; Yosumin!
\\
\addlinespace
\multicolumn{2}{@{}l@{}}{\textcolor{promptGuard}{\textit{Retrieved experience channels $R_k$}}} \\
\addlinespace
\textcolor{channelLt}{\makecell[l]{Preference\\extraction ($\mathrm{pref}$)}} & \textcolor{channelLt}{Strong Valve narrative / puzzle-FPS taste: Half-Life 2 saga and Portal series dominate, with sandbox creativity (Garry's Mod) and occasional indie rogue-likes (Isaac).} \textcolor{channelDis}{Dislikes: Little engagement with pure MOBA test clients; competitive live-service shooters are secondary to story/puzzle FPS.} \\
\addlinespace
\textcolor{channelSt}{\makecell[l]{Context\\extraction ($\mathrm{ctx}$)}} & \textcolor{channelSt}{Recent high-engagement cluster is Portal 2 + Half-Life episodes; short-term focus is completing the Valve narrative loop rather than opening new live-service grinders.} \\
\addlinespace
\textcolor{channelIp}{\makecell[l]{Reasoning\\extraction ($\mathrm{reas}$)}} & \textcolor{channelIp}{Episode Two is the direct narrative continuation of Episode One already in history; CS:GO / Warframe offer multiplayer loops less aligned with the story-FPS preference.} \\
\addlinespace
\textcolor{channelCfCond}{\makecell[l]{Counterfactual\\extraction ($\mathrm{cf}$)}} & \textcolor{channelCfAnchor}{Anchor (Half-Life 2: Episode Two): Continues the exact Half-Life 2 story the user already invested in.} \textcolor{channelCfReject}{Contrast (Warframe): Free-to-play grind / live-service loop; weak narrative continuity with Portal/HL2.} \textcolor{channelCfCond}{If the user wanted a long-horizon multiplayer grind instead of finishing a single-player story arc, the contrast would rank higher.} \\
\addlinespace
\textcolor{promptStruct}{Student output} &
\textcolor{promptStruct}{RANKED LIST:} \textcolor{promptStruct}{1. Half-Life 2: Episode Two}; \textcolor{channelLike}{2. Left 4 Dead 2}; \textcolor{channelLike}{3. Terraria}; 4. PAYDAY 2; 5. Counter-Strike: Global Offensive; 6. Warframe; 7. The Expendabros; 8. Yosumin!
\\[2pt]
& \textcolor{channelIp}{Rationale: Episode Two leads via long-term Valve narrative memory and item-perception continuity; co-op FPS is secondary; mismatched casual / grind titles sink.}
\\
\bottomrule
\end{tabular}
\end{table*}

\section{Failure Cases and Qualitative Memory Edits}
\label{app:failure-cases}


This appendix complements the positive qualitative cases in Section~\ref{sec:results-qualitative} and the end-to-end traces in Appendix~\ref{app:examples} with \emph{failure modes} and \emph{borderline edits}: backbones that lose to GraphRAG, controller edits that rewrite taste too aggressively, and the short-term compression pattern that lowers lexical specificity while remaining actionable.

\paragraph{H.1 Ranking failures vs.\ GraphRAG.} On ML-1M, rEDMRec trails GraphRAG on Llama~3.1~8B (Impv $=-11.1\%$) and GPT~OSS~20B (Impv $=-3.3\%$; Appendix~\ref{app:table1full}). We attribute the Llama failure to weak instruction-following on the concatenated memory prompt (Section~\ref{sec:discussion}), not to an empty bank: the same bank yields positive Impv on stronger students. On Beauty/Steam the same two backbones again show near-zero or negative Impv (Appendix~\ref{app:beauty}--\ref{app:steam}), so the limitation is backbone-dependent rather than dataset-specific.

\begin{table}[pos=t]
\caption{Failure / borderline RQ1 cells (Impv vs.\ second-best; typically GraphRAG). Negative Impv = GraphRAG ahead.}
\label{tbl:app-failure-impv}
\footnotesize
\centering
\begin{tabular}{@{}ll cc@{}}
\toprule
Dataset & Student & Impv (\%) & Note \\
\midrule
ML-1M & Llama~3.1~8B & $-11.1$ & GraphRAG best; weak IF \\
ML-1M & GPT~OSS~20B & $-3.3$ & near-saturated student \\
Beauty & Llama~3.1~8B & $-4.1$ & not significant \\
Beauty & GPT~OSS~20B & $-0.5$ & tie within noise \\
Steam & Llama~3.1~8B & $-7.3$ & GraphRAG best \\
Steam & GPT~OSS~20B & $-2.0$ & GraphRAG best \\
\bottomrule
\end{tabular}
\end{table}

\paragraph{H.2 Qualitative edits: success vs.\ risk.} Table~\ref{tbl:app-failure-edits} contrasts a beneficial long-term rewrite (Case~S1; also Case~A in the main text), a \emph{taste-flip} risk where debate overwrites an earlier sci-fi profile with noir/crime (Case~R1), short-term compression (Case~C1), and a strongly item-grounded item-perception fix (Case~S2).

\begin{table*}[width=\FullWidth,pos=t]
\caption{Qualitative memory edits from the persisted bank (\texttt{experiments/bank\_evolution\_cases.json}). S~=~success-like; R~=~risk / failure mode; C~=~compression.}
\label{tbl:app-failure-edits}
\footnotesize
\begin{tabular}{@{}p{0.08\linewidth} p{0.08\linewidth} p{0.26\linewidth} p{0.26\linewidth} p{0.22\linewidth}@{}}
\toprule
Case & Spec. & Before & After & Interpretation \\
\midrule
\textbf{S1} / $\mathrm{lt}$ & $0.214\to0.479$ & \textit{``Very limited data: the only rated film is a classic action-adventure, suggesting a preference for high-energy, heroic, escapist storytelling with suspense and...''} & \textit{``Long-term: favors mainstream 1990s action buddy-cop films -- franchise sequels, star-driven chemistry, high-energy action with comedic interplay. Upweight these...''} & Hedge -> ranking rule; +spec. \\
\addlinespace
\textbf{R1} / $\mathrm{lt}$ & $0.450\to0.625$ & \textit{``Core taste is classic, lighthearted sci-fi with ensembles and humor, with some room for action SF. Horror is avoided, and the user responds best to accessible,...''} & \textit{``User 17 strongly prefers classic and neo-noir/crime prestige dramas. Downweight short-term popularity signals and upweight niche/indie, 1990s-era, and foreign-...''} & Taste flip risk: sci-fi -> noir; may discard valid prior signal. \\
\addlinespace
\textbf{C1} / $\mathrm{st}$ & $0.300\to0.500$ & \textit{``No recent viewing items were provided, so no emerging short-term interests can be detected.''} & \textit{``Session: prioritize late-80s/90s Hollywood action comedies with buddy dynamics and franchise entries for top slots; include at least one diverse alternative pe...''} & Empty note -> session rule; aggregate st spec. can drop. \\
\addlinespace
\textbf{S2} / $\mathrm{ip}$ & $0.350\to0.850$ & \textit{``The strongest match was only ranked 5th, so Hit@10 was good but Hit@1/MRR suffered. For this user, place the best Rocky-like inspirational drama at rank 1 when...''} & \textit{``Rocky (1976): User 4 chose Rocky over a higher-ranked 80s action title, signaling a preference for 1970s character-driven underdog sports dramas. When ranking,...''} & Vague Hit@1 complaint -> item-keyed boost. \\
\bottomrule
\end{tabular}
\end{table*}

\paragraph{H.3 Capacity-dependent channel reversals.} Channel ablations (Section~\ref{sec:results-h2}) show that removing long-term, item-perception, or counterfactual memory can \emph{improve} HR@1 on the strongest ablation-panel student (\texttt{gpt-5-mini}), i.e.\ the bank can inject noise when the backbone already ranks well from candidates alone. Short-term context remains the only consistently beneficial channel across tiers --- a practical failure mode for ``always retrieve all four channels'' deployments on saturated students.

\end{document}